\documentclass{amsart}

\usepackage{ulem}
\usepackage[english]{babel}
\usepackage[autostyle=true]{csquotes}
\usepackage{amsfonts,amsmath,amsthm,amscd,amssymb,latexsym,mathtools}

\usepackage[hyphens]{url}
\usepackage{hyperref}
\hypersetup{%
    breaklinks=true, %
    colorlinks=true, 
    frenchlinks=true,%
    linkcolor=red, %
    citecolor=red, %
    filecolor=magenta, %
    urlcolor=magenta, %
    linktoc=all,     
}
\usepackage[ruled]{algorithm2e}
\usepackage{graphicx}
\usepackage{float}
\usepackage[final]{microtype} 
\usepackage{longtable}
\usepackage{tabularx, booktabs}
\usepackage{array}
\usepackage{placeins}

\newcommand{\protref}[1]{\hyperref[prot:#1]{protocol~\ref{prot:#1}}}
\newcommand{\algref}[1]{\hyperref[alg:#1]{algorithm~\ref{alg:#1}}}
\newcommand{\secref}[1]{\hyperref[sec:#1]{section~\ref{sec:#1}}}
\newcommand{\subsecref}[1]{\hyperref[subsec:#1]{subsec.~\ref{subsec:#1}}}
\newcommand{\subsubsecref}[1]{\hyperref[subsubsec:#1]{subsubsec.~\ref{subsubsec:#1}}}
\newcommand{\figref}[1]{\hyperref[fig:#1]{fig.~\ref{fig:#1}}}
\newcommand{\tabref}[1]{\hyperref[tab:#1]{table~\ref{tab:#1}}}
\newcommand{\defref}[1]{\hyperref[def:#1]{definition~\ref{def:#1}}}
\newcommand{\thmref}[1]{\hyperref[thm:#1]{theorem~\ref{thm:#1}}}
\newcommand{\lemmaref}[1]{\hyperref[lemma:#1]{lemma~\ref{lemma:#1}}}
\newcommand{\corollaryref}[1]{\hyperref[corollary:#1]{corollary~\ref{corollary:#1}}}
\newcommand{\remarkref}[1]{\hyperref[rmk:#1]{remark~\ref{rmk:#1}}}

\newcommand{\commandof}[2]{\mathmbox{#1\roundbrack{#2}}}
\newcommand{\func}[2]{\commandof{#1}{#2}}
\newcommand{\brackets}[1]{\mathmbox{\left\lbrace#1\right\rbrace}}

\newcommand{\roundbrack}[1]{\mathmbox{\left(#1\right)}}

\newcommand{\floor}[1]{\left\lfloor#1\right\rfloor}

\newcommand{\F}{\mathbb{F}}

\newcommand{\N}{\mathbb{N}}
\newcommand{\K}{\mathbb{K}}

\newcommand{\wt}[1]{\func{\mathrm{wt}}{#1}}

\renewcommand{\email}[1]{\normalsize\href{mailto:#1}{#1}\par}

\newtheorem{theorem}{Theorem}
\numberwithin{theorem}{section} 

\newtheorem{remark}{Remark}
\numberwithin{remark}{section} 
\newtheorem{corollary}[theorem]{Corollary}
\newtheorem{definition}{Definition}
\numberwithin{definition}{section} 

\numberwithin{equation}{section}
\numberwithin{figure}{section}

\newcounter{protocol}
\makeatletter
\newenvironment{protocol}[1][]{%
    \setcounter{algocf}{\value{protocol}}%
    \begin{algorithm}[#1]%
        \refstepcounter{protocol}%
        \hyper@refstepcounter{protocol}%
        \SetAlgorithmName{Protocol}{Protocol \theprotocol}{\theprotocol}
        \small
        \SetInd{0.5em}{0.5em}
    }{\end{algorithm}}
\makeatother

\usepackage{amsaddr}

\title[Decoding algorithm for HL-codes]{Decoding algorithm for HL-codes and performance of the DHH-cryptosystem - a candidate for post-quantum cryptography}

\author{Giuseppe Filippone}
\address{Department of Mathematics and Computer Science, University of Palermo, Via Archirafi 34, 90123 Palermo, Italy}
\email{giuseppe.filippone01@unipa.it}

\author{Carolin Hannusch}
\address{Department of Computer Science, Faculty of Informatics, University of Debrecen, Kassai út 26, 4028 Debrecen, Hungary}
\email{hannusch.carolin@inf.unideb.hu}

\usepackage{cite}

\begin{document}

	\begin{abstract}
		We give a decoding algorithm for a class of error-correcting codes, which can be used in the DHH-cryptosystem, which is a candidate for post-quantum cryptography, since it is of McEliece type. Furthermore, we implement the encryption and decryption algorithms for this cryptosystem and investigate its performance.
	\end{abstract}


    \maketitle

    \section{Preliminaries}

    The well-known asymmetric cryptosystem introduced by McEliece in \cite{mceliece1978public} originally makes use of a general Goppa code, since there is a general decoder for these codes. Although their decoding can be handled easily, the construction of Goppa codes is not easy and implementations for generating Goppa codes usually needs much memory and time. This makes the application of McEliece's cryptosystem not convenient. Nevertheless, McEliece's cryptosystem is a candidate for post-quantum cryptography, as researchers' effort on attacking McEliece's cryptosystem show e.g.~in \cite{bernstein2008attacking}, \cite{repka2014overview}, \cite{nojima2008semantic} and \cite{baldi2016enhanced}.
    Throughout the last decades, researchers worked on decreasing the memory-need of a McEliece cryptosystem \cite{shoufan2009novel}, \cite{roth2020classic} and making it applicaple for small devices \cite{eisenbarth2009microeliece}. Furthermore, researchers also worked on modifying McEliece's cryptosystem (see e.g.~\cite{shrestha2014new}, \cite{ivanov2020new}) in such a way that it would be more convenient for technological applications. One of these is the use of a class of error-correcting codes, called the HL-codes \cite{hannusch2016construction}.
    This cryptosystem was first described in \cite{domosi2019cryptographic} and we refer to at as the DHH-cryptosystem. The main purpose of the current paper is to give a decoding algorithm for the HL-code and to apply this algorithm in the decryption of DHH.
    The introduced decoding algorithm is a variant of Reed's decoding algorithm for Reed-Muller codes \cite{ReedDecoding}. As a byproduct, we rewrite the notations of Reed's paper in the decoding algorithm. Using the decoder, we give a decryption function for the DHH cryptosystem \cite{domosi2019cryptographic}.
    Further, we provide some information about the performance of the DHH cryptosystem using a $(1024, 512, 32)$- HL code and a $(4096, 2048, 64)$ HL-code.

    \section{Construction of the HL-code}\label{sec:2}

    In \cite{hannusch2016construction} a class of error-correcting codes is introduced as ideals in a modular group algebra. These codes were later called HL-codes. Some codes from this family, namely of length $1024$ and of length $4096$ were considered for the use in a McEliece cryptosystem \cite{domosi2019cryptographic}. We will refer to this cryptosystem as the DHH-cryptosystem. In this section, we give a combinatorial construction of an HL-code of length $2^m.$ Note that the constructed code is always a $(2^{2k}, 2^{2k-1}, 2^k)$-code for some positive integer $k.$

    In the following, we give a method for the construction of a generator matrix of an HL-code of length $2^m,$ where $m$ is even, that is $m = 2l$ for some integer $l.$
    The first vectors of the generator matrix are constructed in the following. Let $$v_0 = (\underbrace{1,\ldots,1}_{2^m})$$
    $$v_1 = (0,1,0,1,\ldots, 0,1)$$
    $$v_2 = (0,0,1,1,0,0,1,1 \ldots, 0,0,1,1)$$
    $$\vdots$$
    $$v_{m-1} = (\underbrace{0,\ldots,0}_{2^{m-2}},\underbrace{1,\ldots,1}_{2^{m-2}}, \underbrace{0,\ldots,0}_{2^{m-2}}, \underbrace{1,\ldots,1}_{2^{m-2}})$$
    $$v_m =  (\underbrace{0,\ldots,0}_{2^{m-1}},\underbrace{1,\ldots,1}_{2^{m-1}})$$

    Two vectors are multiplied by multiplying their coordinates. For all positive integers $j<l,$ we consider all possible products of $j$ vectors from the set $\{v_1,\ldots, v_m\},$ i.e.~we construct all vectors of the form

    $$v_{i_1,i_2,\ldots,i_j} = \prod_{i_1,i_2,\ldots,i_j \in \{1,\ldots,m\}} v_{i_1}v_{i_2}\ldots v_{i_j}$$

    At this point, we remark that the generator matrix $$(v_0,v_1,\ldots,v_m,v_{1,2},\ldots,v_{\underbrace{i_1,i_2,\ldots,i_j}_{i_1,i_2,\ldots,i_j \in \{1,\ldots,m\}}})^T$$ generates a Reed-Muller code of length $2^m$ with minimum distance $2^{l+1}.$

    For $j=l$ we consider only those products of $j$ vectors from the set $\{v_1,\ldots, v_m\},$ whose indices build up a maximal complement-free set.

    \begin{definition}
        Let $a=(a_1,\ldots,a_m)$ be a binary vector of length $m.$ Then its complement is the binary vector $\textbf{1}-a = (1-a_1,\ldots, 1-a_m).$ Let $X$ be the set of all binary vectors of length $m=2l,$ where each of these vectors contains exactly $l$ $1$-s. Then $Y$ is called a complement-free set, if it is a subset of $X$ and for each $a\in Y \implies 1-a \not \in Y.$ If $\mid Y\mid = \frac{1}{2} \binom{2l}{l},$ then $Y$ is maximal.
    \end{definition}

    Thus, we take into consideration the following products
     $$v_{i_1,i_2,\ldots,i_j} = \prod_{i_1,i_2,\ldots,i_j \in Y} v_{i_1}v_{i_2}\ldots v_{i_j},$$
    where $Y$ is a maximal complement-free set of the binary vectors of length $2l$ with exactly $l$ $1$-s.
    Then the generator matrix $$(v_0,v_1,\ldots,v_m,v_{1,2},\ldots,v_{\underbrace{i_1,i_2,\ldots,i_j}_{i_1,i_2,\ldots,i_j \in \{1,\ldots,m\}}}, \ldots, v_{\underbrace{i_1,i_2,\ldots,i_j}_{i_1,i_2,\ldots,i_j \in Y}} )^T$$ generates an HL-code of length $2^m$ and dimension $2^{m-1}$ with minimum distance $2^l.$

    \section{Decoder}\label{sec:3}

    \subsection{Decoding algorithm}

        In this section we describe the decoding algorithm \cite{ReedDecoding}
        designed by I. S. Reed. We will use some notations also used in the book \cite{MacWilliamsSloane}
        (chapter $ 13 $) written by F.J. MacWilliamns and N.J.A. Sloane.

        \begin{definition}{(Decoding)}
            Let $ \K $ be a field of cardinality $ q $, and let $ C $ be a $ {[n, k, d]}_q $ a linear
            code defined by a generator matrix $ G \in \K^{k \times n} $, that is,
            $ C = \brackets{v \cdot G \in \K^n \colon v \in \K^k} $.
            For any word $ a \in \K^k $, and $ e \in \K^n $ such that $ \wt{e} \le \floor{\frac{d - 1}{2}} $,
            where $ \wt{e} $ is the Hamming weight of the word $ e $, the decoding of the word
            $ x = a \cdot G + e $ compute the error word $ e $ and returns the codeword $ a \cdot G $ or the word $ a $.
        \end{definition}

        The Reed-Muller decoding algorithm is a \textbf{majority-based binary decoding} which depends on the
        structure of the generator matrix $ G $. We use the generator matrix of an HL-code as it is described in \secref{2}.
        A pairwise product of $ j $ different $ v_i $
        is considered as an element of \textbf{degree} $ j $.

        Every codeword belonging to a linear code may be expressed as a linear combination of the rows of the
        generator matrix. In particular, if $ a = \roundbrack{a_0, \ldots, a_{k - 1}} $,
        then
        \begin{equation*}
            x = a \cdot G = \roundbrack{x_0, \ldots, x_{n - 1}} = \sum\limits_{i = 0}^{k - 1} a_i v_i,
        \end{equation*}
        where, for $ i > m $, $ v_i $ is the pairwise product of the vectors $ v_j $, with $ 1 \le j \le m $.
        Hence, every $ a_i $ is the \textbf{coefficient of degree} $ j $ of $ \binom{m}{j} $ pairwise products of $ v_i $, with $ 1 \le i \le m $.
        From the latter expression, it is possibile to retrieve a set of binary equations,
        known as \textbf{redundancy relations} (see \subsecref{red-rel}), related to each $ a_i $ of the form:
        \begin{equation}\label{eq:eq-set}
            \brackets{\bigoplus\limits_{e \in I_1} x_e, \bigoplus\limits_{e \in I_2} x_e, \ldots, \bigoplus\limits_{e \in I_l} x_e}
        \end{equation}
        where $ \bigoplus $ is the binary sum and, for $ 1 \le s \le l $,
        $ I_s \ne \emptyset $, $ \bigcup\limits_{s = 1}^{l} I_s = \brackets{0, 1, \ldots, n - 1} $,
        and $ I_r \cap I_s = \emptyset $ for every $ r \ne s $.
        Having the set in \eqref{eq:eq-set}, if $ r $ is the number of the equations equals to $ 1 $, then
        \begin{equation*}
            a_i =
            \begin{cases}
                1 \text{ if } r > \frac{l}{2},\\
                0 \text{ if } r < \frac{l}{2}.
            \end{cases}
        \end{equation*}
        Note that if $ r = \frac{l}{2} $, then it is not possible to determine the value of $ a_i $
        (see \thmref{aj-eq-set-bound} and \corollaryref{majority-aj}).

        \begin{theorem}\label{thm:aj-eq-set-bound}
            Let $ a_j $ be a coefficient of degree $ u $, and let
            \begin{equation*}
                S_i = \bigoplus\limits_{e \in I_i} x_e, \qquad 1 \le i \le l,
            \end{equation*}
            be the $ i $-th redundancy relation of $ a_j $, where
            $ I_i \ne \emptyset $, for every $ i $,
            $ \bigcup\limits_{i = 1}^{l} I_i = \brackets{1, \ldots, n} $,
            and $ I_r \cap I_s = \emptyset $ for every $ r \ne s $.

            If $ x = a \cdot G + e $, where $ e $ is an error vector of weight $ t \le \floor{\frac{d - 1}{2}} $,
            then there are at least $ l - t $ redundancy relations for $ a_j $ with the same outcome, that is, $ S_r = S_s $
            for $ r, s \in R \subseteq \brackets{1, \ldots, l} $, with $ \left\lvert R \right\rvert \ge l - t $.
        \end{theorem}
        \begin{proof}
            Since the redundancy relations for each $ a_j $ form a partition of the $ n $ components
            of the vector $ x $, then each the component $ e_i = 1 $ of $ e $ affects only one of the redundancy relations.
            Hence, if the weight of $ e $ is equal to $ 1 $, then only one relation have a different outcome
            with respect the other outcomes.

            If the weight of $ e $ is equal to $ 2 $, then zero or two relations outcomes differ from the other outcomes. Indeed, if
            $ e_r = 1 $ and $ e_s = 1 $ affects the same relation $ S_i $, then in $ S_i $ one has that $ e_r \oplus e_s = 0 $
            and the outcome of $ S_i $ not change. However, if $ e_r = 1 $ and $ e_s = 1 $ affects two different relations,
            then both relations outcomes differ from the other outcomes.
            Hence, if the weight of $ e $ is equal to $ 2 $, then either $ l $ or $ l - 2 $ relations have the same outcomes.

            In general, if the weight of the error vector $ e $ is $ t $, then there are at least $ l - t $ relations
            with the same outcome.
        \end{proof}

        \begin{corollary}\label{corollary:majority-aj}
            Let $ a_j $ be a coefficient of degree $ u $ of $ x = a \cdot G + e $, where $ G \in \F_2^{k \times n} $ is the generator
            matrix of a HL-code, and $ e $ is an arbitrary binary error vector of weight $ t \le \floor{\frac{d - 1}{2}} $.
            Let $ l $ be the number of redundancy relations for $ a_j $, and let $ r $ be the number of redundancy relations
            for $ a_j $ whose outcomes are equal to $ 1 $.

            If $ r > \frac{l}{2} $, then $ a_j = 1 $, while if $ r < \frac{l}{2} $,
            then $ a_j = 0 $.
        \end{corollary}
        \begin{proof}
            Since $ m $ is an even number, then $ t \le \floor{\frac{d - 1}{2}} = \floor{\frac{2^{\frac{m}{2}} - 1}{2}} = 2^{\frac{m}{2} - 1} - 1 $.
            Moreover, it easy to check that $ l = \frac{n}{2^u} = \frac{2^m}{2^u} = 2^{m - u} $ (see \eqref{eq:delta-op}),
            with $ 1 \le u \le \frac{m}{2} $.
            Hence, we have that $ 2^{\frac{m}{2}} \le l \le 2^{m - 1} $, and $ t < \frac{l}{2} $.
            Therefore, from \thmref{aj-eq-set-bound}, there are $ l - t > \frac{l}{2} $ redundancy relations with the same
            outcome $ b \in \brackets{0, 1} $, and $ a_j $ is equal to $ b $.
        \end{proof}

        The decoding process starts with the highest-degree coefficients $ a_i $
        and ends with the lowest-degree ones in a iterative way. These decoding method also works
        for an HL-code.

        More precisely, since the elements of highest-degree in a HL-code have degree equals to $ \frac{m}{2} $
        (the elements from the complement-free set defined in \secref{2}),
        one has to decode these latter elements from $ x = a \cdot G + e $,
        and remove them from $ x $ to apply again the decoding
        to
        \begin{equation*}
            x^\prime = x - \sum\limits_{i \in Y\text{-set}} a_i v_i,
        \end{equation*}
        where  $ Y $-set is the complement-free set of indices associated to the coefficients $ a_i $ of degree $ \frac{m}{2} $.
        Since, from \corollaryref{majority-aj}, it is possible to determine the proper value of $ a_i $, then
        one has to repeat this process by finding the next highest-degree coefficients $ a_i $,
        and remove them from $ x^\prime $. At the end of this process, one is left with the vector $ x^\prime = a_0 v_0 + e $.
        However, since $ v_0 = \roundbrack{1, \ldots, 1} $, in order to find $ a_0 $, one has to count the number $ r $ of $ 1 $ in
        $ x^\prime $. Since $ x^\prime \in {\F_2}^n $,
        \begin{equation*}
            a_0 =
            \begin{cases}
                1 \text{ if } r > \frac{n}{2},\\
                0 \text{ if } r < \frac{n}{2}.
            \end{cases}
        \end{equation*}
        Note that again if $ r = \frac{n}{2} $, then one cannot find $ a_0 $ and decode $ x $ (see \thmref{a0}).
        Finally, $ e = x^\prime - a_0 v_0 $ and $ a \cdot G = x - e $.

        \begin{theorem}\label{thm:a0}
            Let $ a = \roundbrack{a_0, a_1, \ldots, a_{k - 1}} $ be an arbitrary binary vector,
            and let $ x = a \cdot G + e $ be the codeword modified
            by an arbitrary binary vector $ e $ such that $ \wt{e} \le \floor{\frac{d - 1}{2}} $, where $ G \in \F_2^{k \times n} $
            is the generator matrix of a HL-code.

            If $ \wt{a_0 v_0 + e} > \frac{n}{2} $, then $ a_0 = 1 $, while if $ \wt{a_0 v_0 + e} < \frac{n}{2} $,
            then $ a_0 = 0 $.
        \end{theorem}
        \begin{proof}
            From \corollaryref{majority-aj}, it is possible to determine each value of $ a_j $, for $ j = 1, \ldots, k - 1 $,
            by using the redundancy relations of $ a_j $. Thus, one may compute
            $ x^\prime = x - \sum\limits_{i = 1}^{k - 1} a_i v_i = a_0 v_0 + e \in \F_2^n $.
            Since $ t \le 2^{\frac{m}{2} - 1} - 1 $ and $ \frac{n}{2} = 2^{m - 1} $, then one has that
            $ t < \frac{n}{2} $. Therefore, if $ a_0 = 0 $, then $ x^\prime = e $ and $ \wt{x^\prime} = t < \frac{n}{2} $,
            while if $ a_0 = 1 $, then $ x^\prime = v_0 + e $ and $ \wt{x^\prime} = n - t > \frac{n}{2} $.
        \end{proof}

        \begin{remark}
            Note that \corollaryref{majority-aj} and \thmref{a0} prove that the majority rule properly determine each coefficient $ a_j $,
            for $ j = 0, \ldots, k - 1 $.
        \end{remark}

    \subsection{Compute the redundancy relations}\label{subsec:red-rel}

        In this section we present the operator $ \Delta $ used by Reed to compute the redundancy relations
        for every coefficient $ a_i $, for $ 0 \le i < k $.

        In order to define the operator $ \Delta $, we define in the following a function able to compute which vectors $ v_j $,
        with $ 1 \le j \le m $, we have to pairwise multiply with the coefficient $ a_i $ in
        $ x = a \cdot G = \sum\limits_{i = 0}^{k - 1} a_i v_i $.

        First, let $ M $ be the set of integer $ \brackets{1, \ldots, m} $, and
        let $ \binom{M}{j} $ be a $ j $-combination of the elements in $ M $ such that $ e \in \binom{M}{j} $
        is a tuple $ e $ of $ j $ indices in the form $ e = (l_1, \ldots, l_j) $. Also, we define a natural ordering for
        $ \binom{M}{j} $ such that, for any $ (e_1, e_2) \in \binom{M}{j} \times \binom{M}{j} $,
        \begin{equation}\label{eq:sort-elem}
            (l_{1, 1}, \ldots, l_{1, j}) = e_1 \le e_2 = (l_{2, 1}, \ldots, l_{2, j}) \iff l_{1, s} \le l_{2, s} \text{ for all } 1 \le s \le j.
        \end{equation}
        Moreover, we refer to the $ i $-th element $ e_i $
        of $ \binom{M}{j} $ as the element such that
        \begin{equation*}
             e_i \ge e_{i - 1} \ge \ldots \ge e_1,
        \end{equation*}
        where $ e_1 $ is the minimum element in $ \binom{M}{j} $.
        Note that, the complement-free set $ Y $, defined in \secref{2},
        is a subset of $ \binom{M}{m / 2} $. However, for the complement-free set $ Y $,
        we put $ e_i \in \mathmbox{Y\text{-set}} $ as the $ i $-th element inserted into
        $ \mathmbox{Y\text{-set}} $ (as this set is randomly generated for security purposes).
        Hence, we consider the order of $ e \in \mathmbox{Y\text{-set}} $
        as a list with the policy \textbf{FIFO} (First In First Out), that is, $ e_0 \in \mathmbox{Y\text{-set}} $
        is the first element inserted into $ \mathmbox{Y\text{-set}} $, and $ e_t \in \mathmbox{Y\text{-set}} $,
        with $ t = \frac{1}{2} \binom{m}{m / 2} - 1 $, is the last element inserted into $ \mathmbox{Y\text{-set}} $.

        For $ i \in \N $, let $ \nu $ be the function defined as
        \begin{equation}\label{eq:nu}
            \func{\nu}{i} = \underset{t \in \N}{\max} \roundbrack{i \ge \sum\limits_{j = 0}^{t} \binom{m}{j}},
        \end{equation}
        and let $ \lambda $ be the function defined as
        \begin{equation}\label{eq:lambda}
            \func{\lambda}{i} = i - \sum\limits_{j = 0}^{\func{\nu}{i}} \binom{m}{j}.
        \end{equation}

        Hence, we have a correspondence between the $ i $-th row of $ G $ and the indices of the pairwise
        product of vectors $ v_j $, with $ 1 \le j \le m $, given by the following function:
        \begin{equation}\label{eq:f-i}
            f(i) =
            \begin{cases}
                e_{\func{\lambda}{i}} \in \binom{M}{\func{\nu}{i} + 1} &\qquad \mathmbox{\text{if}} \quad 1 \le i < k - \frac{1}{2} \binom{m}{m / 2},\\
                e_{\func{\lambda}{i}} \in \mathmbox{Y\text{-set}} &\qquad \mathmbox{\text{if}} \quad k - \frac{1}{2} \binom{m}{m / 2} \le i < k,
            \end{cases}
        \end{equation}
        where $ e_{\func{\lambda}{i}} \in \mathmbox{Y\text{-set}} $ is the $ \func{\lambda}{i} $-th element of
        $ \mathmbox{Y\text{-set}} $. Note that, for $ 1 \le i \le m $, $ f(i) = e_i \in \binom{M}{1} = (i) $.

        For instance, if $ m = 6 $ and $ i = 7 $, then $ \func{\nu}{7} = 1 $ and $ e_{\func{\lambda}{7}} = (1, 2) $ is the first
        element of $ \binom{M}{2} $. Also, if $ i = 28 $, then $ \func{\nu}{28} = 2 $ and $ e_{\func{\lambda}{28}} $
        is equal to the sixth element of the complement-free set $ Y \subset \binom{M}{3} $.

        Hence, we may rewrite $ x = a \cdot G $ as
        \begin{equation*}
            \sum\limits_{i = 0}^{k - 1} a_i \roundbrack{\prod\limits_{j \in f(i)} v_j},
        \end{equation*}
        where $ \prod $ is the pairwise product.

        Each redundancy relation related to a coefficient $ a_i $ may be determined by a recursive relation.
        In order to describe this latter, we define the function
        \begin{equation}\label{eq:psi}
            \begin{array}{rrl}
                \psi \colon &\N \times \N &\longrightarrow \brackets{0, 1}\\
                &\roundbrack{i, k} &\longmapsto i_{k - 1}\\
            \end{array},
        \end{equation}
        where $ i_{k - 1} $ is the $ k $-th LSB (least significant bit)
        of the binary representation of the integer $ i $, that is,
        $ i = {\roundbrack{i_{n - 1}, i_{n - 2}, \ldots, i_k, i_{k - 1}, \ldots, i_1, i_0}}_2 $,
        and the function
        \begin{equation}\label{eq:phi}
            \begin{array}{rrl}
                \phi \colon &\N \times \N &\longrightarrow \N\\
                    &\roundbrack{i, k} &\longmapsto i + {(-1)}^{\func{\psi}{i, k}} \cdot 2^{k - 1}\\
            \end{array}.
        \end{equation}
        Hence, $ \func{\phi}{i, k} $ change the $ k $-th LSB of the integer $ i $. If
        $ \func{\psi}{i, k} $ is equal to $ 0 $, then $ \func{\phi}{i, k} = i + 2^{k - 1} $, while if
        $ \func{\psi}{i, k} $ is equal to $ 1 $, then $ \func{\phi}{i, k} = i - 2^{k - 1} $.

        We also extend the function $ \phi $ as follow:
        \begin{equation}
            \begin{array}{rrl}
            \Phi \colon &\N \times \N^l &\longrightarrow \N\\
                &\roundbrack{i, (k_1, \ldots, k_l)} &\longmapsto i + \sum\limits_{j = 1}^{l} {(-1)}^{\func{\psi}{i, k_j}} \cdot 2^{k_j - 1}\\
            \end{array}.
        \end{equation}
        Therefore, the function $ \func{\Phi}{i, (k_1, \ldots, k_l)} $ changes at the same time
        the $ k_s $-th LSB of $ i $, with $ s = 1, \ldots, l $.

        The recursive relation to find the redundancy relations is given by:
        \begin{equation}\label{eq:delta-op}
            \begin{array}{rlrrr}
                \underset{j}{\Delta}\ \ x_i &= &x_i &\oplus &x_{\phi(i, j)},\\
                \underset{(j_1, \ldots, j_t)}{\Delta}\ \ x_i &= &\underset{(j_1, \ldots, j_{t - 1})}{\Delta}\ \ x_i &\oplus &\underset{(j_1, \ldots, j_{t - 1})}{\Delta}\ \ x_{\phi(i, j_t)},\\
            \end{array}
        \end{equation}
        for $ i = 0, \ldots, k - 1 $, and $ \roundbrack{j_1, \ldots, j_t} \in \binom{M}{t} $.

        More precisely, if $ J_t = \brackets{j_1, \ldots, j_t} $, then applying the recursion we get that
        \begin{equation} \label{eq:delta-expansion}
            \underset{(j_1, \ldots, j_t)}{\Delta}\ \ x_i = x_i \bigoplus\limits_{r = 1}^{t} \roundbrack{
                \bigoplus\limits_{e \in \binom{J_t}{r}} x_{\func{\Phi}{i, e}}}.
        \end{equation}

        Note that, in this case, the order of the elements belonging to $ \binom{J_t}{r} $ does not matter.

        \bigskip

        The operator $ \Delta $ computes the redundancy relations for each coefficient $ a_j $ in
        $ x = a \cdot G $, with $ G $ the generator matrix of the code.
        In particular, for $ i = 0, \ldots, k - 1 $, if $ 1 \le j \le m $, then one
        computes the redundancy relations for the coefficients $ a_j $ with $ \underset{j}{\Delta}\ \ x_i $,
        while  if $ m < j < k $, then one computes the redundancy relations for
        the coefficient $ a_j $ with $ \underset{(j_1, \ldots, j_t)}{\Delta}\ \ x_i $,
        where $ \roundbrack{j_1, \ldots, j_t} \in \binom{M}{t} $.

    \subsection{Little example}

        In this section we show how to find the redundancy relations for each coefficient $ a_i $ without the $ \Delta $,
        and subsequently we show how to simply find them with $ \Delta $.

        Let $ m = 4 $, so $ n = 2^m = 16 $, and $ k = 2^{m - 1} = 8 $. Also,
        suppose for simplicity that $ Y $-set is equal to $ \roundbrack{(1, 4), (1, 3), (1, 2)} $ in this order.
        In this case, we have that
        \begin{equation*}
            G = \left(\begin{matrix}
                v_0 \\
                v_1 \\
                v_2 \\
                v_3 \\
                v_4 \\
                v_1 v_4 \\
                v_1 v_3 \\
                v_1 v_2 \\
            \end{matrix} \right) = \qquad
            \begin{pmatrix}
                1 & 1 & 1 & 1 & 1 & 1 & 1 & 1 & 1 & 1 & 1 & 1 & 1 & 1 & 1 & 1\\
                0 & 1 & 0 & 1 & 0 & 1 & 0 & 1 & 0 & 1 & 0 & 1 & 0 & 1 & 0 & 1\\
                0 & 0 & 1 & 1 & 0 & 0 & 1 & 1 & 0 & 0 & 1 & 1 & 0 & 0 & 1 & 1\\
                0 & 0 & 0 & 0 & 1 & 1 & 1 & 1 & 0 & 0 & 0 & 0 & 1 & 1 & 1 & 1\\
                0 & 0 & 0 & 0 & 0 & 0 & 0 & 0 & 1 & 1 & 1 & 1 & 1 & 1 & 1 & 1\\
                0 & 0 & 0 & 0 & 0 & 0 & 0 & 0 & 0 & 1 & 0 & 1 & 0 & 1 & 0 & 1\\
                0 & 0 & 0 & 0 & 0 & 1 & 0 & 1 & 0 & 0 & 0 & 0 & 0 & 1 & 0 & 1\\
                0 & 0 & 0 & 1 & 0 & 0 & 0 & 1 & 0 & 0 & 0 & 1 & 0 & 0 & 0 & 1\\
            \end{pmatrix}.
        \end{equation*}
        Hence, $ a \cdot G = a_0 v_0 + a_1 v_1 + a_2 v_2 + a_3 v_3 + a_4 v_4 + a_5 v_1 v_4 + a_6 v_1 v_3 + a_7 v_1 v_2 $.
        In order to determine each $ a_i $, one does not takes into account the error vector $ e $, e.g. one takes $ x = a \cdot G $
        and not $ x = a \cdot G + e $, as the majority rule for the redundancy relations will give
        the proper coefficient $ a_i $. Therefore, the components $ x_i $ of the vector
        $ x = a \cdot G $ are

        \begin{equation}\label{eq:x-i-eq}
            \begin{aligned}[c]
                x_0 &= a_0, \\
                x_1 &= a_0 \oplus a_1, \\
                x_2 &= a_0 \oplus a_2, \\
                x_3 &= a_0 \oplus a_1 \oplus a_2 \oplus a_7, \\
                x_4 &= a_0 \oplus a_3, \\
                x_5 &= a_0 \oplus a_1 \oplus a_3 \oplus a_6, \\
                x_6 &= a_0 \oplus a_2 \oplus a_3, \\
                x_7 &= a_0 \oplus a_1 \oplus a_2 \oplus a_3 \oplus a_6 \oplus a_7,
            \end{aligned} \qquad
            \begin{aligned}[c]
            	x_8 &= a_0 \oplus a_4,\\
                x_9 &= a_0 \oplus a_1 \oplus a_4 \oplus a_5,\\
                x_{10} &= a_0 \oplus a_2 \oplus a_4,\\
                x_{11} &= a_0 \oplus a_1 \oplus a_2 \oplus a_4 \oplus a_5 \oplus a_7,\\
                x_{12} &= a_0 \oplus a_3 \oplus a_4,\\
                x_{13} &= a_0 \oplus a_1 \oplus a_3 \oplus a_4 \oplus a_5 \oplus a_6,\\
                x_{14} &= a_0 \oplus a_2 \oplus a_3 \oplus a_4,\\
                x_{15} &= a_0 \oplus a_1 \oplus a_2 \oplus a_3 \oplus a_4 \oplus a_5 \oplus a_6 \oplus a_7.
            \end{aligned}
        \end{equation}
        First, since one has to start with the highest-degree coefficients,
        we determine the redundancy relation of the $ a_5 $, $ a_6 $, $ a_7 $ from \eqref{eq:x-i-eq}
        \begin{equation}\label{eq:rec-rel-1}
            \begin{aligned}[c]
                &\begin{aligned}[c]
                    a_5 &= x_0 \oplus x_1 \oplus x_8 \oplus x_9 = \\
                    &= x_2 \oplus x_3 \oplus x_{10} \oplus x_{11} = \\
                    &= x_4 \oplus x_5 \oplus x_{12} \oplus x_{13} = \\
                    &= x_6 \oplus x_7 \oplus x_{14} \oplus x_{15},
                \end{aligned} \qquad
                \begin{aligned}[c]
                    a_6 &= x_0 \oplus x_1 \oplus x_4 \oplus x_5 = \\
                    &= x_2 \oplus x_3 \oplus x_6 \oplus x_7 = \\
                    &= x_8 \oplus x_9 \oplus x_{12} \oplus x_{13} = \\
                    &= x_{10} \oplus x_{11} \oplus x_{14} \oplus x_{15},
                \end{aligned}\\&{}\\
                &\begin{aligned}[c]
                    a_7 &= x_0 \oplus x_1 \oplus x_2 \oplus x_3 = \\
                    &= x_4 \oplus x_5 \oplus x_6 \oplus x_7 = \\
                    &= x_8 \oplus x_9 \oplus x_{10} \oplus x_{11} = \\
                    &= x_{12} \oplus x_{13} \oplus x_{14} \oplus x_{15}.
                \end{aligned}
            \end{aligned}
        \end{equation}
        In order to determine the next highest-degree coefficient,
        we have to consider $ x^\prime = x - a_5 v_1 v_4 - a_6 v_1 v_3 - a_7 v_1 v_2 $.
        Note that, for $ x^\prime $, we have that $ a_5 = a_6 = a_7 = 0 $, so we may remove them from \eqref{eq:x-i-eq}
        and we find
        \begin{equation}\label{eq:x-i-eq-reduced}
            \begin{aligned}[c]
                x^\prime_0 &= a_0, \\
                x^\prime_1 &= a_0 \oplus a_1, \\
                x^\prime_2 &= a_0 \oplus a_2, \\
                x^\prime_3 &= a_0 \oplus a_1 \oplus a_2, \\
                x^\prime_4 &= a_0 \oplus a_3, \\
                x^\prime_5 &= a_0 \oplus a_1 \oplus a_3, \\
                x^\prime_6 &= a_0 \oplus a_2 \oplus a_3, \\
                x^\prime_7 &= a_0 \oplus a_1 \oplus a_2 \oplus a_3,
            \end{aligned} \qquad
            \begin{aligned}[c]
                x^\prime_8 &= a_0 \oplus a_4,\\
                x^\prime_9 &= a_0 \oplus a_1 \oplus a_4,\\
                x^\prime_{10} &= a_0 \oplus a_2 \oplus a_4,\\
                x^\prime_{11} &= a_0 \oplus a_1 \oplus a_2 \oplus a_4,\\
                x^\prime_{12} &= a_0 \oplus a_3 \oplus a_4,\\
                x^\prime_{13} &= a_0 \oplus a_1 \oplus a_3 \oplus a_4,\\
                x^\prime_{14} &= a_0 \oplus a_2 \oplus a_3 \oplus a_4,\\
                x^\prime_{15} &= a_0 \oplus a_1 \oplus a_2 \oplus a_3 \oplus a_4.
            \end{aligned}
        \end{equation}
        Now, we determine the redundancy relation of the $ a_1 $, $ a_2 $, $ a_3 $ and $ a_4 $ from \eqref{eq:x-i-eq-reduced}:
        \begin{equation}\label{eq:rec-rel-2}
            \begin{array}{lllll}
                a_1 &= x^\prime_0 \oplus x^\prime_1 &= x^\prime_2 \oplus x^\prime_3 &= x^\prime_4 \oplus x^\prime_5 &= x^\prime_6 \oplus x^\prime_7 = \\
                    &= x^\prime_8 \oplus x^\prime_9 &= x^\prime_{10} \oplus x^\prime_{11} &= x^\prime_{12} \oplus x^\prime_{13} &= x^\prime_{14} \oplus x^\prime_{15}, \\\\
                a_2 &= x^\prime_0 \oplus x^\prime_2 &= x^\prime_1 \oplus x^\prime_3 &= x^\prime_4 \oplus x^\prime_6 &= x^\prime_5 \oplus x^\prime_7 = \\
                    &= x^\prime_8 \oplus x^\prime_{10} &= x^\prime_9 \oplus x^\prime_{11} &= x^\prime_{12} \oplus x^\prime_{14} &= x^\prime_{13} \oplus x^\prime_{15}, \\\\
                a_3 &= x^\prime_0 \oplus x^\prime_4 &= x^\prime_1 \oplus x^\prime_5 &= x^\prime_2 \oplus x^\prime_6 &= x^\prime_3 \oplus x^\prime_7 = \\
                    &= x^\prime_8 \oplus x^\prime_{12} &= x^\prime_9 \oplus x^\prime_{13} &= x^\prime_{10} \oplus x^\prime_{14} &= x^\prime_{11} \oplus x^\prime_{15}, \\\\
                a_4 &= x^\prime_0 \oplus x^\prime_8 &= x^\prime_1 \oplus x^\prime_9 &= x^\prime_2 \oplus x^\prime_{10} &= x^\prime_3 \oplus x^\prime_{11} = \\
                    &= x^\prime_4 \oplus x^\prime_{12} &= x^\prime_5 \oplus x^\prime_{13} &= x^\prime_6 \oplus x_{14} &= x_7 \oplus x_{15}. \\
            \end{array}
        \end{equation}
        Hence, from \eqref{eq:rec-rel-1} on $ x $ and from \eqref{eq:rec-rel-2} on $ x^\prime $, we determine
        \begin{equation*}
            x^{\prime\prime} = x - a_1 v_1 - a_2 v_2 - a_3 v_3 - a_4 v_4 - a_5 v_1 v_4 - a_6 v_1 v_3 - a_7 v_1 v_2 = a_0 v_0.
        \end{equation*}
        Recall that if $ x = a \cdot G + e $, then $ x^{\prime\prime} = a_0 v_0 + e $ (see \thmref{a0}).

        Now, we apply the operator $ \Delta $ to find the redundancy relations. For instance, we directly
        find the first relation of $ a_1 $ as
        \begin{equation*}
            \begin{aligned}
                \underset{1}{\Delta}\  x_0 = x_0 \oplus x_{\func{\phi}{0, 1}} = x_0 \oplus x_1.
            \end{aligned}
        \end{equation*}
        Recall that as each relation form a partition (see \eqref{eq:eq-set}) of $ \brackets{x_0, \ldots, x_{15}} $. Therefore, the next relation
        is given by
        \begin{equation*}
            \begin{aligned}
                \underset{1}{\Delta}\  x_2 = x_2 \oplus x_{\func{\phi}{2, 1}} = x_2 \oplus x_3.
            \end{aligned}
        \end{equation*}
        Hence, after the computation of each redundancy relation, one knows the next $ x_i $ to take
        into account. Note that to compute $ \underset{1}{\Delta}\  x_1 $ is time consuming as
        \begin{equation*}
            \underset{1}{\Delta}\  x_1 = x_1 \oplus x_{\func{\phi}{1, 1}} = x_1 \oplus x_0 = \underset{1}{\Delta}\  x_0.
        \end{equation*}
        To summarize, we have that the redundancy relations for each $ a_i $ are given by
        \begin{equation}
            \begin{aligned}
                a_1 &= \underset{\func{f}{1}}{\Delta}\  x_0 = \underset{\func{f}{1}}{\Delta}\  x_2 = \underset{\func{f}{1}}{\Delta}\  x_4 = \underset{\func{f}{1}}{\Delta}\  x_6 = \\
                    &= \underset{\func{f}{1}}{\Delta}\  x_8 = \underset{\func{f}{1}}{\Delta}\  x_{10} = \underset{\func{f}{1}}{\Delta}\  x_{12} = \underset{\func{f}{1}}{\Delta}\  x_{14}, \\\\
                a_2 &= \underset{\func{f}{2}}{\Delta}\  x_0 = \underset{\func{f}{2}}{\Delta}\  x_1 = \underset{\func{f}{2}}{\Delta}\  x_4 = \underset{\func{f}{2}}{\Delta}\  x_5 = \\
                    &= \underset{\func{f}{2}}{\Delta}\  x_8 = \underset{\func{f}{2}}{\Delta}\  x_{9} = \underset{\func{f}{2}}{\Delta}\  x_{12} = \underset{\func{f}{2}}{\Delta}\  x_{13}, \\\\
                a_3 &= \underset{\func{f}{3}}{\Delta}\  x_0 = \underset{\func{f}{3}}{\Delta}\  x_1 = \underset{\func{f}{3}}{\Delta}\  x_2 = \underset{\func{f}{3}}{\Delta}\  x_3 = \\
                    &= \underset{\func{f}{3}}{\Delta}\  x_8 = \underset{\func{f}{3}}{\Delta}\  x_9 = \underset{\func{f}{3}}{\Delta}\  x_{10} = \underset{\func{f}{3}}{\Delta}\  x_{11}, \\\\
                a_4 &= \underset{\func{f}{4}}{\Delta}\  x_0 = \underset{\func{f}{4}}{\Delta}\  x_1 = \underset{\func{f}{4}}{\Delta}\  x_2 = \underset{\func{f}{4}}{\Delta}\  x_3 = \\
                    &= \underset{\func{f}{4}}{\Delta}\  x_4 = \underset{\func{f}{4}}{\Delta}\  x_5 = \underset{\func{f}{4}}{\Delta}\  x_6 = \underset{\func{f}{4}}{\Delta}\  x_7, \\\\
                a_5 &= \underset{\func{f}{5}}{\Delta}\  x_0 = \underset{\func{f}{5}}{\Delta}\  x_2 = \underset{\func{f}{5}}{\Delta}\  x_4 = \underset{\func{f}{5}}{\Delta}\  x_6, \\\\
                a_6 &= \underset{\func{f}{6}}{\Delta}\  x_0 = \underset{\func{f}{6}}{\Delta}\  x_2 = \underset{\func{f}{6}}{\Delta}\  x_8 = \underset{\func{f}{6}}{\Delta}\  x_{10}, \\\\
                a_7 &= \underset{\func{f}{7}}{\Delta}\  x_0 = \underset{\func{f}{7}}{\Delta}\  x_4 = \underset{\func{f}{7}}{\Delta}\  x_8 = \underset{\func{f}{7}}{\Delta}\  x_{12}. \\
            \end{aligned}
        \end{equation}
        Note that, for instance, the first redundancy relation for $ a_6  $ is $ \underset{\func{f}{6}}{\Delta}\  x_0 $ and
        since $ f(6) = e_{\func{\lambda}{6}} \in \mathmbox{Y\text{-set}} = e_2 \in \mathmbox{Y\text{-set}} = (1, 3) $, we have that
        \begin{equation*}
            \begin{aligned}
                \underset{\func{f}{6}}{\Delta}\  x_0 &= \underset{(1, 3)}{\Delta}\  x_0 = \underset{1}{\Delta}\  x_0 \oplus \underset{1}{\Delta}\  x_{\func{\phi}{0, 3}} = \\
                                                   &= \roundbrack{x_0 \oplus x_{\func{\phi}{0, 1}}} \oplus \roundbrack{x_{\func{\phi}{0, 3}} \oplus x_{\func{\Phi}{0, (1, 3)}}} = \\
                                                   &= x_0 \oplus x_1 \oplus x_4 \oplus x_5 = \\
                                                   &= x_0 \oplus \roundbrack{\bigoplus\limits_{r = 1}^{2} \roundbrack{
                                                       \bigoplus\limits_{e \in \binom{(1, 3)}{r}} x_{\func{\Phi}{0, e}}}} =
                                                   x_0 \oplus x_{\func{\Phi}{0, (1)}} \oplus x_{\func{\Phi}{0, (3)}} \oplus x_{\func{\Phi}{0, (1, 3)}},
            \end{aligned}
        \end{equation*}
        where the last equivalence comes from the expansion of the operator $ \Delta $ in \eqref{eq:delta-expansion}.
        Note that, by using the operator $ \Delta $, it is not necessary to find the redundancy relations as
        in \eqref{eq:x-i-eq} and \eqref{eq:x-i-eq-reduced}.

    \section{Implementations and computing results}\label{sec:4}

    In this section we show the pseudocodes to compute the generator matrix $ G $ in \secref{2},
    all the redundancy relations for each $ a_j $, and decoding a vector $ x = a \cdot G + e $, with
    $ e $ a vector error such that $ \wt{e} = \floor{\frac{d - 1}{2}} $, with $ d $ as the minimum distance
    of the HL-code.
    In the end of this section we show the results of our implementation for an HL-code with $ m = 10 $,
    and $ m = 12 $.

    First, we defined in \algref{psi-phi-func} a function which computes
    \begin{equation*}
        \func{\phi}{i, k} = i + {(-1)}^{\func{\psi}{i, k}} \cdot 2^{k - 1},
    \end{equation*}
    where $ \psi $ and $ \phi $ are the functions defined in \eqref{eq:psi} and \eqref{eq:phi},
    respectively.

   \FloatBarrier
        \normalem
        \begin{algorithm}[tbh!!]
            \label{alg:psi-phi-func}
            \caption{Change-Bit}
            \small
            \SetAlgoNoLine
            \SetAlgoNoEnd
            \SetInd{0.5em}{0.5em}
            \SetFillComment
            \DontPrintSemicolon

            \SetStartEndCondition{ $ \left(\right. $}{$ \left. \right) $}{$ \left. \right) $}\SetAlgoBlockMarkers{}{\}}%
            \SetKwIF{If}{ElseIf}{Else}{if}{ \{}{elif}{else \{}{}%
            \SetKwProg{Fn}{Function}{ \{}{}
            \AlgoDisplayBlockMarkers%

            \SetKwData{i}{i}
            \SetKwData{k}{k}
            \SetKwFunction{ChangeBit}{Change-Bit}

            \Fn(\tcp*[h]{this function computes $ \func{\phi}{\i, \k} $}){\ChangeBit{\i, \k}}{%

                \KwData{$ \i \in \N $, $ \k \in \N $}
                \KwResult{$ \i \pm 2^{\k - 1} $}
                \BlankLine

                $ {(j_{n - 1}, j_{n - 2}, \ldots, j_1, j_0)}_2 \longleftarrow \i $\tcp*{get the binary representation of \i}
                \If(\tcp*[h]{$ j_{k - 1} == 1 $}){$ j_{k - 1} = 1 $}{%
                    \KwRet{$ \i - 2^{\k - 1} $}%
                }
                \Else(\tcp*[h]{$ j_{k - 1} == 0 $}){%
                    \KwRet{$ \i + 2^{\k - 1} $}%
                }
            }
        \end{algorithm}%
        \ULforem%

    Next, we defined \algref{nu-lambda-func} to compute the functions $ \lambda $,
    as in \eqref{eq:lambda}, and $ \nu $, as in \eqref{eq:nu}. More precisely, this algorithm
    returns the tuple $ (\func{\lambda}{i}, \func{\nu}{i}) $ for some integer $ i \in \brackets{0, \ldots, k - 1} $.

      \FloatBarrier
        \normalem
        \begin{algorithm}[tbh!!]
            \label{alg:nu-lambda-func}
            \caption{Row-To-Comb-Index}
            \small
            \SetAlgoNoLine
            \SetAlgoNoEnd
            \SetInd{0.5em}{0.5em}
            \SetFillComment
            \DontPrintSemicolon

            \SetStartEndCondition{ $ \left(\right. $}{$ \left. \right) $}{$ \left. \right) $}\SetAlgoBlockMarkers{}{\}}%
            \SetKwFor{While}{while}{ \{}{}%
            \SetKwProg{Fn}{Function}{ \{}{}
            \AlgoDisplayBlockMarkers%

            \SetKwData{i}{i}
            \SetKwData{mm}{m}
            \SetKwData{t}{t}
            \SetKwData{s}{s}
            \SetKwData{sp}{$ s^\prime $}
            \SetKwData{ip}{$ i^\prime $}
            \SetKwFunction{RowCombIdx}{Row-To-Comb-Index}

            \Fn(\tcp*[h]{this function computes $ \func{\lambda}{\i} $}){\RowCombIdx{\i, \mm}}{%
                \KwData{$ \i \in \brackets{0, \ldots, k - 1} $, $ \mm \in \N $, where $ \i $ is
                    the row of the generator matrix $ G \in {\F_2}^{k \times n} $,
                    $ k = 2^{\mm - 1} $, and $ n = 2^{\mm} $}
                \KwResult{$ (\ip, \t) $ such that $ e_{\ip} \in \binom{M}{\t + 1} $ or $ e_{\ip} \in \mathmbox{Y\text{-set}} $}
                \BlankLine
                $ \t \longleftarrow -1 $\;
                $ \s \longleftarrow 0 $\;
                $ \sp \longleftarrow 0 $\;
                \While{$ \i \ge \s $}{%
                    $ \sp \longleftarrow \s $\;
                    $ \t \longleftarrow \t + 1 $\;
                    $ \s \longleftarrow \s + \binom{\mm}{\t} $\;
                }
                \tcp{$ \t - 1 $ is the maximum integer such that $ \i \ge \sp = \sum\limits_{j = 0}^{\t - 1} \binom{\mm}{j} $}
                \KwRet{$ (\i - \sp, \t - 1) $}
            }
        \end{algorithm}%
        \ULforem%

\newpage
    In \algref{f-i-func} we defined the function to compute $ \func{f}{i} $ as in \eqref{eq:f-i}. In
    particular, we return the tuple $ (i) $, if $ 1 \le i \le m $, the $ \func{\lambda}{i} $-th element of
    $ \binom{\brackets{1, \ldots, m}}{\func{\nu}{i} + 1} $ (sorted with the rule in \eqref{eq:sort-elem}),
    if $ m < i < k - \frac{1}{2}\binom{m}{m / 2} $,
    and the $ \func{\lambda}{i} $-th element of the complement-free set $ Y $, if $ k - \frac{1}{2}\binom{m}{m / 2} \le i < k $.

   \FloatBarrier
        \normalem
        \begin{algorithm}[tbh!]
            \label{alg:f-i-func}
            \caption{Row-To-Comb}
            \small
            \SetAlgoNoLine
            \SetAlgoNoEnd
            \SetInd{0.5em}{0.5em}
            \SetFillComment
            \DontPrintSemicolon

            \SetStartEndCondition{ $ \left(\right. $}{$ \left. \right) $}{$ \left. \right) $}\SetAlgoBlockMarkers{}{\}}%
            \SetKwIF{If}{ElseIf}{Else}{if}{ \{}{elif}{else \{}{}%
            \SetKwProg{Fn}{Function}{ \{}{}
            \AlgoDisplayBlockMarkers%

            \SetKwData{i}{i}
            \SetKwData{mm}{m}
            \SetKwData{t}{t}
            \SetKwData{k}{k}
            \SetKwData{Yset}{$ Y $-set}
            \SetKwData{kp}{$ k^\prime $}
            \SetKwData{ip}{$ i^\prime $}
            \SetKwFunction{RowComb}{Row-To-Comb}
            \SetKwFunction{RowCombIdx}{Row-To-Comb-Index}

            \Fn(\tcp*[h]{this function computes $ \func{f}{\i} $}){\RowComb{\i, \mm, \Yset}}{%
                \KwData{$ \i \in \brackets{1, \ldots, k - 1} $, $ \mm \in \N $, where $ \i $ is
                    the row of the generator matrix $ G \in {\F_2}^{k \times n} $,
                    $ k = 2^{\mm - 1} $, and $ n = 2^{\mm} $}
                \KwResult{$ \func{f}{i} $, where $ f $ is the funcion in \eqref{eq:f-i}}
                \BlankLine
                \If{$ 1 \le \i \le \mm $}{
                    \KwRet{$ (\i) $}
                }
                $ \kp \longleftarrow \k - \frac{1}{2}\binom{\mm}{\mm / 2} $\;
                $ (\ip, \t) \longleftarrow $ \RowCombIdx{\i, \mm}\;
                \If{$ \mm < \i < \kp $}{
                    \KwRet{$ e_{\ip} \in \binom{\brackets{1, \ldots, \mm}}{\t + 1} $}
                }
                \Else{
                    \KwRet{$ e_{\ip} \in \Yset $}
                }
            }
        \end{algorithm}%
        \ULforem%

    In \algref{redundancy-relation} we compute the operator $ \underset{e}{\Delta}\  x_i $, with $ e \in \N^l $,
    for a coefficient $ a_j $. We use the recursive definition in \eqref{eq:delta-op} of this operator because is more
    efficient than the iterative ones in \eqref{eq:delta-expansion}.
    Indeed, in the recursive definition is easy to define a cache to put the already
    computed results and, therefore, avoid to recompute them. For instance,
    \begin{equation*}
        \begin{aligned}
            \underset{(1, 2, 3, 4)}{\Delta}\  x_0 &= \underset{(1, 2, 3)}{\Delta}\  x_0 \oplus \underset{(1, 2, 3)}{\Delta}\  x_8 =\\
                                                &= \roundbrack{\underset{(1, 2)}{\Delta}\  x_0 \oplus \underset{(1, 2)}{\Delta}\  x_4} \oplus \roundbrack{\underset{(1,2)}{\Delta}\  x_8 \oplus \underset{(1, 2)}{\Delta}\  x_{12}} = \\
                                                &= \roundbrack{\underset{1}{\Delta}\  x_0 \oplus \underset{1}{\Delta}\  x_2 \oplus \underset{1}{\Delta}\  x_4 \oplus \underset{1}{\Delta}\  x_6} \oplus
                                                \roundbrack{\underset{1}{\Delta}\  x_8 \oplus \underset{1}{\Delta}\  x_{10} \oplus
                                                    \underset{1}{\Delta}\  x_{12} \oplus \underset{1}{\Delta}\  x_{14}}.
        \end{aligned}
    \end{equation*}
    Hence, starting from $ a_1 $ to $ a_{k - 1} $, one may cache all the inner results of the operator $ \Delta $ in order to speedup the
    next computations. In the above case, if we cached the results of $ \underset{(1, 2, 3)}{\Delta}\  x_0 $ during a previous computation,
    then we need to compute only half of the elements in $ \underset{(1, 2, 3, 4)}{\Delta}\  x_0 $. Note that it is not easy to
    define a caching method with the iterative definition of $ \Delta $.

   \FloatBarrier
        \normalem
        \begin{algorithm}[tbh!]
            \label{alg:redundancy-relation}
            \caption{Redundancy-Relation}
            \small
            \SetAlgoNoLine
            \SetAlgoNoEnd
            \SetInd{0.5em}{0.5em}
            \SetFillComment
            \DontPrintSemicolon

            \SetStartEndCondition{ $ \left(\right. $}{$ \left. \right) $}{$ \left. \right) $}\SetAlgoBlockMarkers{}{\}}%
            \SetKwIF{If}{ElseIf}{Else}{if}{ \{}{elif}{else \{}{}%
            \SetKwProg{Fn}{Function}{ \{}{}
            \AlgoDisplayBlockMarkers%

            \SetKwData{e}{e}
            \SetKwData{ep}{$ e^\prime $}
            \SetKwData{i}{i}
            \SetKwData{ip}{$ i^\prime $}
            \SetKwData{C}{C}
            \SetKwData{t}{t}
            \SetKwData{Ru}{$ R_1 $}
            \SetKwData{Rd}{$ R_2 $}
            \SetKwData{R}{R}
            \SetKwFunction{ChangeBit}{Change-Bit}
            \SetKwFunction{RedundancyRelation}{Redundancy-Relation}

        \Fn(\tcp*[h]{\scriptsize compute $ \underset{\e}{\Delta}\  x_i $ for $ a_j $, with $ \func{f}{j} = \e $}){\RedundancyRelation{\e, \i, \C}}{%

            \KwData{$ (j_1, \ldots, j_t) = \e = \func{f}{j} $ for some $ j $, $ i \in \brackets{1, \ldots, n} $, $ \C $,
                where $ n = 2^m $, and $ \C $ is a dictionary used to cache the already computed redundancy relations}
            \KwResult{the set of indices of the components of $ x $ in $ \underset{\e}{\Delta}\  x_i $}
            \BlankLine

            \If{$ \t = 0 $}{\KwRet{$ (\emptyset, \C) $}\;}
            \If(\tcp*[h]{check if the pair $ (\e, \i) $ was already computed}){(\e, \i) is in \C}{%
                \tcp{return the cached value $ \func{\C}{\e, \i} = \func{\C}{(j_1, \ldots, j_t), \i} $ and the cache \C}
                \KwRet{$ (\func{\C}{\e, \i}, \C) $}%
            }
            \If(\tcp*[h]{if \t == $ 1 $, then return $ \underset{j_1}{\Delta}\  x_i = x_i \oplus x_{\func{\phi}{i, j_1}} $ and \C}){\t = 1}{%
                $ \ip \longleftarrow $ \ChangeBit{\i, $ j_1 $}\;
                $ \R \longleftarrow \brackets{i, \ip} $\tcp*{set of indices $ i $ and $ \ip $}%
                \KwRet{$ (\R, \C) $}%
            }\Else(\tcp*[h]{if $ \t > 1 $, then compute $ \underset{\e}{\Delta}\  x_i $ by recursion}){%
                $ \ep \longleftarrow (j_1, \ldots, j_{t - 1}) $\;
                $ \ip \longleftarrow $ \ChangeBit{\i, $ j_t $}\;
                $ (\Ru, \C) \longleftarrow $ \RedundancyRelation{\ep, \i, \C} \tcp*{compute $ \underset{\ep}{\Delta}\  x_i $}
                $ \func{\C}{\ep, \i} \longleftarrow \Ru $ \tcp*{update cache with $ \underset{\ep}{\Delta}\  x_i $}
                $ (\Rd, C) \longleftarrow $ \RedundancyRelation{\ep, \ip, \C} \tcp*{compute $ \underset{\ep}{\Delta}\  x_{\ip} $}
                $ \R \longleftarrow \Ru \cup \Rd $\;
                $ \func{\C}{\ep, \ip} \longleftarrow \Rd $ \tcp*{update cache with $ \underset{\ep}{\Delta}\  x_{i^\prime} $}
                $ \func{\C}{\e, \i} \longleftarrow \R $ \tcp*{update cache with $ \underset{\e}{\Delta}\  x_i $}
                \tcp{\scriptsize return the set of indices of the components of $ x $ in $ \underset{\e}{\Delta}\  x_i $ and the cache $ C $}
                \KwRet{$ (\R, \C) $}%
            }
        }
        \end{algorithm}%
        \ULforem%

In order to speedup the \algref{aj-redundancy-relations} (explained below), we return $ \underset{e}{\Delta}\  x_i $ as the set of indices
of the components of $ x $ taken into account. For instance,
$ \underset{(1, 2, 3, 4)}{\Delta}\  x_0 = \bigoplus\limits_{i = 0}^{15} x_i $ will be represented as the set $ \brackets{0, \ldots, 15} $.

    In \algref{aj-redundancy-relations} we compute the set of all redundancy relations for $ a_j $ as in \eqref{eq:eq-set}.
    As the redundancy relations of $ a_j $ form a partition of the $ n $ components of the vector $ x = a \cdot G + e $,
    in order to compute $ \underset{e}{\Delta}\  x_i $ for the next component $ x_i $ of $ x $,
    we define a set of the indices $ I_n = \brackets{0, \ldots, n - 1} $ and, for each computed redundancy relation,
    we remove the components $ x_j $ taken into account in the previous relation.
    For instance, $ \underset{(1, 2, 3, 4)}{\Delta}\  x_0 = \bigoplus\limits_{i = 0}^{15} x_i $, therefore
    we remove the indices $ \brackets{0, \ldots, 15} $ from $ I_n $.

   \FloatBarrier
        \normalem
        \begin{algorithm}[tbh!]
            \label{alg:aj-redundancy-relations}
            \caption{Redundancy-Relations}
            \small
            \SetAlgoNoLine
            \SetAlgoNoEnd
            \SetInd{0.5em}{0.5em}
            \SetFillComment
            \DontPrintSemicolon

            \SetStartEndCondition{ $ \left(\right. $}{$ \left. \right) $}{$ \left. \right) $}\SetAlgoBlockMarkers{}{\}}%
            \SetKwIF{If}{ElseIf}{Else}{if}{ \{}{elif}{else \{}{}%
            \SetKwFor{While}{while}{ \{}{}%
            \SetKwProg{Fn}{Function}{ \{}{}
            \AlgoDisplayBlockMarkers%

            \SetKwData{e}{e}
            \SetKwData{i}{i}
            \SetKwData{j}{j}
            \SetKwData{mm}{m}
            \SetKwData{n}{n}
            \SetKwData{C}{C}
            \SetKwData{R}{R}
            \SetKwData{Rc}{$ R_c $}
            \SetKwData{In}{$ I_n $}
            \SetKwData{Yset}{$ Y $-set}
            \SetKwFunction{RowComb}{Row-To-Comb}
            \SetKwFunction{RedundancyRelations}{Redundancy-Relations}
            \SetKwFunction{RedundancyRelation}{Redundancy-Relation}

            \Fn(\tcp*[h]{\scriptsize compute the set of all redundancy relations for $ a_j $}){\RedundancyRelations{\j, \mm, \Yset, \C}}{%
                \KwData{$ \j \in \brackets{1, \ldots, k - 1} $, $ \mm \in \N $, $ \Yset $, $ \C $, where $ k = 2^{\mm - 1} $,
                and $ \C $ is a dictionary used to cache the already computed redundancy relations}
                \KwResult{$ \brackets{\underset{\e}{\Delta}\  x_i} $ for $ i = 0, \ldots, n - 1 $, $ \e $ and the updated cache $ \C $}
                \BlankLine

                $ \n \longleftarrow 2^{\mm} $\;
                $ \R \longleftarrow $ empty list\;
                $ \In \longleftarrow \{0, \ldots, n - 1\} $\;
                $ \e \longleftarrow $ \RowComb{\j, \mm, \Yset}\;
                \While{\In is not empty}{
                    $ \i \longleftarrow $ lowest element in $ \In $\;
                    $ (\Rc, \C) \longleftarrow \RedundancyRelation{\e, \i, \C} $\;
                    append $ \Rc $ to $ \R $\;
                    $ \In \longleftarrow \In \setminus \Rc $\tcp*{remove the set of indices in \Rc from \In}
                }%
                \KwRet{(\R, \e, \C)}
            }
        \end{algorithm}%
        \ULforem%

    In \algref{decoding-set} we compute all the redundancy relations for each $ a_j $,
    with $ j = j_{\text{start}}, \ldots, j_{\text{end}} $, and we put them into a dictionary
    indexed by the coefficient degree of $ a_j $. In order to explain the former choice,
    note that by definition the first $ k - \frac{1}{2}\binom{m}{m / 2} $ rows of the generator matrix
    $ G $ are fixed, that is, two different generator matrices will have the same first $ k - \frac{1}{2}\binom{m}{m / 2} $
    rows. Hence, to speedup the whole process, we compute $ a_j $, for $ 0 \le j < k - \frac{1}{2}\binom{m}{m / 2} $, and save the cache
    and the redundancy relations into a file. As the last $ \frac{1}{2}\binom{m}{m / 2} $ rows of the generator matrix
    are random and depend on the complement-free set $ Y $, during the decoding we may compute the remaining redundancy relations,
    that is, the redundancy relations of $ a_j $, for $ k - \frac{1}{2}\binom{m}{m / 2} \le j < k $. Hence, we run the algorithm
    for $ j_{\text{start}} = 0 $ and $ j_{\text{end}} = k - \frac{1}{2}\binom{m}{m / 2} - 1 $ to save the results into a file,
    and during the key-generation we run the algorithm for $ j_{\text{start}} = k - \frac{1}{2}\binom{m}{m / 2} $ and $ j_{\text{end}} = k - 1 $.

   \FloatBarrier
        \normalem
        \begin{algorithm}[tbh!]
            \label{alg:decoding-set}
            \caption{Redundancy-Relations-Set}
            \small
            \SetAlgoNoLine
            \SetAlgoNoEnd
            \SetInd{0.5em}{0.5em}
            \SetFillComment
            \DontPrintSemicolon

            \SetStartEndCondition{ $ \left(\right. $}{$ \left. \right) $}{$ \left. \right) $}\SetAlgoBlockMarkers{}{\}}%
            \SetKwIF{If}{ElseIf}{Else}{if}{ \{}{elif}{else \{}{}%
            \SetKwFor{For}{for}{ \{}{}%
            \SetKwProg{Fn}{Function}{ \{}{}
            \AlgoDisplayBlockMarkers%

            \SetKwData{e}{e}
            \SetKwData{j}{j}
            \SetKwData{mm}{m}
            \SetKwData{C}{C}
            \SetKwData{l}{r}
            \SetKwData{jstart}{$ j_{\text{start}} $}
            \SetKwData{jend}{$ j_{\text{end}} $}
            \SetKwData{R}{R}
            \SetKwData{Rc}{$ R_c $}
            \SetKwData{Yset}{$ Y $-set}
            \SetKwFunction{Length}{Length}
            \SetKwFunction{CFreeset}{Complement-Free-Set}
            \SetKwFunction{DecodingSet}{Redundancy-Relations-Set}
            \SetKwFunction{RedundancyRelations}{Redundancy-Relations}

            \Fn(\tcp*[h]{\scriptsize compute the set of all redundancy relations for all $ a_j $, for $ j = \jstart, \ldots, \jend $}){\DecodingSet{\mm, \Yset, \C, \jstart, \jend}}{%
                \KwData{$ \C $ is the cache, $ \mm \in \N $, $ \Yset $, $ \jstart \in \N $, $ \jend \in \N $,
                    with $ 0 \le \jstart < \jend < k $, and $ \Yset $ is the complement-free set from \CFreeset}
                \KwResult{$ \brackets{\underset{\e}{\Delta}\  x_i} $ for each $ a_j $, with $ i = 0, \ldots, 2^{\mm} - 1 $,
                $ j = \jstart, \ldots, \jend $}
                \BlankLine

                $ \R \longleftarrow $ empty dictionary\;
                \For{$ \j = \jstart $ \KwTo $ \jend $}{
                    $ (\Rc, \e, \C) \longleftarrow \RedundancyRelations{\j, \mm, \Yset, \C} $\;
                    $ \l \longleftarrow $ \Length{\e} \tcp*{$ \l $ is the degree of the coefficient $ a_j $}
                    \If{$ \R $ has not key $ \l $}{%
                        $ \func{\R}{\l} \longleftarrow $ empty list\;
                    }
                    insert $ (\j, \Rc) $ into $ \func{\R}{\l} $ \tcp*{insert $ (\j, \Rc) $ into $ \R $ with key $ \l $}
                }%
                \KwRet{\R}
            }
        \end{algorithm}%
        \ULforem%

   \newpage
    In \algref{build-y-set} we compute the complement-free set $ Y $. More precisely, we list all the element
    of $ \binom{\brackets{1, \ldots, m}}{m / 2} $ and we sort them using the rule in \eqref{eq:sort-elem}. We call
    this list as the $ X $-set. It is easy to see that the first half of the previous set is the complement of the second half in reverse order.
    In particular, if $ A $ is the first half of the $ X $-set and $ B $ is the second half in reverse order of the $ X $-set,
    then $ A_i $ is the complement of $ B_i $, where $ A_i $ and $ B_i $ are the $ i $-th element of $ A $ and $ B $, respectively.
    Hence, we may compute the complement-free set $ Y $ by randomly choosing an element in $ A $ or in $ B $.
    In order to add some randomness, we shuffle the two lists in the same way.

    \FloatBarrier
    \normalem
    \begin{algorithm}[tbh!]
        \label{alg:build-y-set}
        \caption{Complement-Free-Set}
        \small
        \SetAlgoNoLine
        \SetAlgoNoEnd
        \SetInd{0.5em}{0.5em}
        \SetFillComment
        \DontPrintSemicolon

        \SetStartEndCondition{ $ \left(\right. $}{$ \left. \right) $}{$ \left. \right) $}\SetAlgoBlockMarkers{}{\}}%
        \SetKwIF{If}{ElseIf}{Else}{if}{ \{}{elif}{else \{}{}%
        \SetKwFor{For}{for}{ \{}{}%
        \SetKwProg{Fn}{Function}{ \{}{}
        \AlgoDisplayBlockMarkers%

        \SetKwData{mm}{m}
        \SetKwData{e}{e}
        \SetKwData{j}{j}
        \SetKwData{r}{r}
        \SetKwData{t}{t}
        \SetKwData{A}{A}
        \SetKwData{B}{B}
        \SetKwData{Mset}{M}
        \SetKwData{Xset}{$ \mathmbox{X\text{-set}} $}
        \SetKwData{Yset}{$ \mathmbox{Y\text{-set}} $}
        \SetKwFunction{CFreeset}{Complement-Free-Set}

        \Fn(\tcp*[h]{build the partial generator matrix a the HL-code of parameter \mm}){\CFreeset{\mm}}{%
            \KwData{$ \mm \in \N $, where $ \mm \mid 2 $}
            \KwResult{The complement-free $ \Yset $}
            \BlankLine

            $ \Yset \longleftarrow $ empty list\;
            $ \Mset \longleftarrow \brackets{0, \ldots, \mm - 1}$ \;
            \tcp{compute all $ \frac{\mm}{2} $-combination of elements in $ \Mset $}
            $ \Xset \longleftarrow $ sorted list of $ \e \in \binom{\Mset}{\mm / 2} $\;
            \tcp{if we take all $ \e $ in $ \binom{\Mset}{\mm / 2} $ such that $ {\e}_1 \le {\e}_2 $, then the second half of $ \Xset $ is the complement
                of its first half}
            $ \t \longleftarrow \binom{\mm}{\mm / 4} $ \tcp*{half of the size of \Xset}
            $ \A \longleftarrow $ first \t elements of \Xset\;
            $ \B \longleftarrow $ last \t elements of \Xset\;
            $ \B \longleftarrow $ reverse $ \B $ \tcp*{$ B = \roundbrack{{\e}_{\t + 1}, \ldots, {\e}_{2 \t}} \mapsto \roundbrack{{\e}_{2 \t}, \ldots, {\e}_{\t + 1}} $}
            $ \A \longleftarrow $ shuffle $ \A $\;
            \tcp{shuffle $ \B $ such that $ {\B}_j $ is always the complement of $ {\A}_j $}
            $ \B \longleftarrow $ shuffle $ \B $ as $ \A $\;
            \For{$ \j = 1 $ \KwTo \t}{
                $ \r \longleftarrow $ random binary number \tcp*{$ \r \in \brackets{0, 1} $ random}
                \If{$ \r = 0 $}{ %
                    append $ {\A}_{\j} $ to $ \Yset $ \tcp*{append the $ \j $-th element of $ \A $ to $ \Yset $}
                }\Else(\tcp*[h]{$ \r == 1 $}){%
                    append $ {\B}_{\j} $ to $ \Yset $ \tcp*{append the $ \j $-th element of $ \B $ to $ \Yset $}
                }
            }
            \KwRet{\Yset}
        }
    \end{algorithm}%
    \ULforem%

   \newpage
    In \algref{create-vi-vector} we define a function to compute the vector $ v_i $ in \secref{2} by using the
    exponentiation by squaring. More precisely, if $ x = (\mathbf{0} | \mathbf{1}) $, with $ \mathbf{0}, \mathbf{1} \in {\F_2}^{2^i} $,
    then $ v_i = x^{\frac{n}{2^{i + 1}}} $. For instance, if $ n = 8 $ and $ x = (0 | 1) $,
    then $ i = 0 $ and $ v_1 = x^{\frac{8}{2}} = {(0 | 1)}^4 = \roundbrack{0, 1, 0, 1, 0, 1, 0, 1} $.

   \FloatBarrier
        \normalem
        \begin{algorithm}[tbh!]
            \label{alg:create-vi-vector}
            \caption{Build-V-Vector}
            \small
            \SetAlgoNoLine
            \SetAlgoNoEnd
            \SetInd{0.5em}{0.5em}
            \SetFillComment
            \DontPrintSemicolon

            \SetStartEndCondition{ $ \left(\right. $}{$ \left. \right) $}{$ \left. \right) $}\SetAlgoBlockMarkers{}{\}}%
            \SetKwIF{If}{ElseIf}{Else}{if}{ \{}{elif}{else \{}{}%
            \SetKwFor{While}{while}{ \{}{}%
            \SetKwProg{Fn}{Function}{ \{}{}
            \AlgoDisplayBlockMarkers%

            \SetKwData{n}{n}
            \SetKwData{i}{i}
            \SetKwData{v}{v}
            \SetKwData{vp}{$ v^\prime $}
            \SetKwData{s}{s}
            \SetKwData{z}{z}
            \SetKwData{u}{u}
            \SetKwFunction{VVector}{Build-V-Vector}

            \Fn{\VVector{\i, \n}}{%
                \KwData{$ \i \in \N $, $ \n \in \N $, with $ \n \mid 2^{\i + 1} $}
                \KwResult{$ \roundbrack{\mathbf{0} | \mathbf{1} | \ldots | \mathbf{0} | \mathbf{1}} = \v \in {\F_2}^{\n} $, where
                $ \mathbf{0}, \mathbf{1} \in {\F_2}^{2^{\i}} $}
                \BlankLine

                $ \s \longleftarrow \dfrac{\n}{2^{\i + 1}} $\;
                $ \vp \longleftarrow $ empty vector\;
                $ \z \longleftarrow \mathbf{0} \in {\F_2}^{2^{\i}} $\;
                $ \u \longleftarrow \mathbf{1} \in {\F_2}^{2^{\i}} $\;
                $ \v \longleftarrow \roundbrack{\z | \u}$\;
                \While{$ \s > 1 $}{
                    \If{\s is odd}{
                        $ \vp \longleftarrow \roundbrack{\vp | \v} $ \tcp*{concatenate \vp and \v}
                        $ \s \longleftarrow \s - 1 $\;
                    }
                    $ \v \longleftarrow \roundbrack{\v | \v} $ \tcp*{concatenate \v with itself}
                    $ \s \longleftarrow \dfrac{\s}{2} $\;
                }
                $ \v \longleftarrow \roundbrack{\v | \vp} $ \tcp*{$ \v = {\roundbrack{\mathbf{0} | \mathbf{1}}}^{\frac{\n}{2^{\i + 1}}} $}
                \KwRet{\v}
            }
        \end{algorithm}%
        \ULforem%

\newpage

    In \algref{build-rm-matrix} we compute the first $ k - \frac{1}{2}\binom{m}{m / 2} $ rows of the generator matrix $ G $.
    In particular, we compute the first $ m + 1 $ rows by using the \algref{create-vi-vector}, and next we
    compute the remaining $ k - \frac{1}{2}\binom{m}{m / 2} - (m + 1) $ rows by computing all the pairwise products
    of the vectors $ v_i $, with $ i = 1, \ldots, m $, taking all the combinations of indices $ \binom{\brackets{1, \ldots, m}}{j} $
    (sorted with the rule in \eqref{eq:sort-elem}), for $ j = 2, \ldots, \frac{m}{2} - 1 $. As these rows will not
    change, we save this matrix into a file. We may read the matrix from the saved file to complete the remaining rows
    of the generator matrix.

   \FloatBarrier
        \normalem
        \begin{algorithm}[tbh!]
            \label{alg:build-rm-matrix}
            \caption{Build-HL-Partial-Generator-Matrix}
            \small
            \SetAlgoNoLine
            \SetAlgoNoEnd
            \SetInd{0.5em}{0.5em}
            \SetFillComment
            \DontPrintSemicolon

            \SetStartEndCondition{ $ \left(\right. $}{$ \left. \right) $}{$ \left. \right) $}\SetAlgoBlockMarkers{}{\}}%
            \SetKwIF{If}{ElseIf}{Else}{if}{ \{}{elif}{else \{}{}%
            \SetKwFor{For}{for}{ \{}{}%
            \SetKwProg{Fn}{Function}{ \{}{}
            \AlgoDisplayBlockMarkers%

            \SetKwData{mm}{m}
            \SetKwData{n}{n}
            \SetKwData{k}{k}
            \SetKwData{i}{i}
            \SetKwData{j}{j}
            \SetKwData{t}{t}
            \SetKwData{G}{G}
            \SetKwData{e}{e}
            \SetKwData{Mset}{M}
            \SetKwFunction{VVector}{Build-V-Vector}
            \SetKwFunction{HLPMatrix}{Build-HL-Partial-Gen-Matrix}

            \Fn(\tcp*[h]{build the partial generator matrix a the HL-code of parameter \mm}){\HLPMatrix{\mm}}{%
                \KwData{$ \mm \in \N $, where $ \mm \mid 2 $}
                \KwResult{The partial generator matrix $ \G \in {\F_2}^{\k \times \n} $ of an HL-code, where $ n = 2^{\mm} $,
                    and $ \k = 2^{\mm - 1} $}
                \BlankLine
                $ \n \longleftarrow 2^{\mm} $\;
                $ \k \longleftarrow 2^{\mm - 1} $\;
                $ \G \longleftarrow $ empty binary matrix of dimension $ \k \times \n $\;
                $ \G_0 \longleftarrow \mathbf{1} \in {\F_2}^{\n} $\;
                \For{$ \i = 1 $ \KwTo $ \mm $}{
                    $ {\G}_{\i} \longleftarrow $ \VVector{$ \i - 1 $, \n} \tcp*{set the $ \i $-th row of $ \G $ as $ {\roundbrack{\mathbf{0} | \mathbf{1}}}^{\frac{\n}{2^{\i}}} $}
                }%
                $ \t \longleftarrow \mm + 1 $\;
                $ \Mset \longleftarrow \brackets{0, \ldots, \mm - 1}$ \;
                \tcp{compute all the pairwise products of the rows at $ \j $ of $ \G $, with $ \j = 1, \ldots, \mm $}
                \For{$ \i = 2 $ \KwTo $ \frac{\mm}{2} - 1 $}{ %
                    \tcp{get all the ordered combinations in $ \binom{\Mset}{\i} $}
                    \ForAll{$ \e $ in $ \binom{\Mset}{\i} $}{%
                        $ {\G}_{\t} \longleftarrow \prod\limits_{\j \in \e} {\G}_{\j} $ \tcp*{pairwise products of rows}
                        $ \t \longleftarrow \t + 1 $;
                    }
                }
                \KwRet{\G}
            }
        \end{algorithm}%
        \ULforem%

\newpage
    In \algref{build-hl-matrix} we compute the complete generator matrix of the HL-code by using the complement-free
    set $ Y $ computed in \algref{build-y-set}, and the partial generator matrix computed using the \algref{build-rm-matrix}.

       \FloatBarrier
        \normalem
        \begin{algorithm}[tbh!]
            \label{alg:build-hl-matrix}
            \caption{Build-HL-Generator-Matrix}
            \small
            \SetAlgoNoLine
            \SetAlgoNoEnd
            \SetInd{0.5em}{0.5em}
            \SetFillComment
            \DontPrintSemicolon

            \SetStartEndCondition{ $ \left(\right. $}{$ \left. \right) $}{$ \left. \right) $}\SetAlgoBlockMarkers{}{\}}%
            \SetKwIF{If}{ElseIf}{Else}{if}{ \{}{elif}{else \{}{}%
            \SetKwFor{For}{for}{ \{}{}%
            \SetKwProg{Fn}{Function}{ \{}{}
            \AlgoDisplayBlockMarkers%

            \SetKwData{mm}{m}
            \SetKwData{e}{e}
            \SetKwData{i}{i}
            \SetKwData{j}{j}
            \SetKwData{G}{G}
            \SetKwData{Yset}{$ \mathmbox{Y\text{-set}} $}
            \SetKwFunction{HLMatrix}{Build-HL-Gen-Matrix}
            \SetKwFunction{CFreeset}{Complement-Free-Set}
            \SetKwFunction{HLPMatrix}{Build-HL-Partial-Gen-Matrix}

            \Fn(\tcp*[h]{build the full generator matrix a the HL-code of parameter \mm}){\HLMatrix{\mm, \G, \Yset}}{%
                \KwData{$ \mm \in \N $, $ \G \in {\F_2}^{k \times n} $,
                    where $ \mm \mid 2 $, $ n = 2^{\mm} $, $ k = 2^{\mm - 1} $,
                    $ \G $ is the partial generator matrix from \HLPMatrix, and
                    $ \Yset $ is the complement-free set from \CFreeset}
                \KwResult{The full generator matrix $ \G \in {\F_2}^{k \times n} $ of an HL-code}
                \BlankLine
                \tcp{compute the index of the first row of $ \G $ to insert the pairwise product from $ \Yset $}
                $ \i \longleftarrow \sum\limits_{j = 0}^{\frac{\mm}{2} - 1} \binom{\mm}{j} + 1 $\;
                \ForAll{$ \e $ in $ \Yset $}{%
                    $ {\G}_{\i} \longleftarrow \prod\limits_{\j \in \e} {\G}_{\j} $ \tcp*{pairwise products of rows}
                    $ \i \longleftarrow \i + 1 $;
                }
                \KwRet{\G}
           }
       \end{algorithm}%
       \ULforem%

\newpage
    In order to decode the word $ x = a \cdot G + e $, we need to apply a multilevel decoding starting
    with the coefficient $ a_j $ of highest-degree. In \algref{decode-error-level} we decode the word $ x $
    for the coefficient of degree $ r $ computed with the \algref{decoding-set}. We also set the proper
    $ a_j $ to the vector $ a^\prime $ that, in the end of \algref{decoding-set}, will be equal to the vector $ a $.

    Note that, it is not necessary to count exactly how many redundancy relations for $ a_j $ are equal to $ 1 $.
    Indeed, let $ s $ be the number of redundancy relations for $ a_j $,
    if the number of ones (or the number of zeros) is greater than $ \frac{s}{2} $,
    then the value of $ a_j $ is equal $ 1 $ or $ 0 $, respectively.

   \FloatBarrier
        \normalem
        \begin{algorithm}[tbh!]
            \label{alg:decode-error-level}
            \caption{Decode-Level-R}
            \small
            \SetAlgoNoLine
            \SetAlgoNoEnd
            \SetInd{0.5em}{0.5em}
            \SetFillComment
            \DontPrintSemicolon

            \SetStartEndCondition{ $ \left(\right. $}{$ \left. \right) $}{$ \left. \right) $}\SetAlgoBlockMarkers{}{\}}%
            \SetKwIF{If}{ElseIf}{Else}{if}{ \{}{elif}{else \{}{}%
            \SetKwFor{For}{for}{ \{}{}%
            \SetKwFor{While}{while}{ \{}{}%
            \SetKwProg{Fn}{Function}{ \{}{}
            \AlgoDisplayBlockMarkers%

            \SetKwData{a}{$ a^\prime $}
            \SetKwData{e}{e}
            \SetKwData{v}{v}
            \SetKwData{x}{x}
            \SetKwData{l}{r}
            \SetKwData{k}{k}
            \SetKwData{n}{n}
            \SetKwData{i}{i}
            \SetKwData{j}{j}
            \SetKwData{t}{t}
            \SetKwData{s}{s}
            \SetKwData{u}{u}
            \SetKwData{z}{z}
            \SetKwData{G}{G}
            \SetKwData{rsum}{h}
            \SetKwData{R}{R}
            \SetKwData{Rj}{$ R_j $}
            \SetKwFunction{Length}{Length}
            \SetKwFunction{DecodingSet}{Redundancy-Relations-Set}
            \SetKwFunction{DecodeRelation}{Decode-Level-R}

            \Fn{\DecodeRelation{\R, \l, \x, \G}}{%
                \KwData{$ \R $, $ \x \in {\F_2}^{\n} $, $ \G \in {\F_2}^{\k \times \n} $, where $ \x $ is the vector
                    to decode, $ \G $ is the generator matrix of the HL-code, and $ \R $ is the dictionary of the redundancy relations
                    for the coefficients $ a_j $ of degree $ \l $ as in \DecodingSet}
                \KwResult{$ \a $ and $ \v $}
                \BlankLine

                $ \k \longleftarrow $ get the number of rows of $ \G $\;%
                $ \n \longleftarrow $ get the number of columns of $ \G $\;%
                $ \a \longleftarrow \mathbf{0} \in {\F_2}^{\k}$\;%
                $ \v \longleftarrow \mathbf{0} \in {\F_2}^{\n}$\;%
                \ForAll{$ (\j, \Rj) $ in $ \func{\R}{\l} $}{%
                    $ \t \longleftarrow 0 $\;%
                    $ \u \longleftarrow 0 $ \tcp*{counter for the $ 1 $}%
                    $ \z \longleftarrow 0 $ \tcp*{counter for the $ 0 $}%
                    $ \s \longleftarrow $ \Length{\Rj} \tcp*{number of rendundancy relations for $ a_j $}%
                    \tcp{count how many $ \e \in \Rj $ are equal to $ 1 $, and how many are equal to $ 0 $}
                    \While{$ \u \le \frac{\s}{2} $ and $ \z \le \frac{\s}{2} $}{%
                        \tcp{$ \rsum $ is sum the components of $ \x $ according to the $ \t $-th redundancy relation $ {\Rj}^{(\t)} $ of $ a_j $}%
                        $ \rsum \longleftarrow \bigoplus\limits_{\e \in {\Rj}^{(\t)}} \bigoplus\limits_{\i \in \e} {\x}_{\i} $\;%
                        \If{$ \rsum = 1 $}{ %
                            $ \u \longleftarrow \u + 1 $\;%
                        }
                        \Else{%
                            $ \z \longleftarrow \z + 1 $\;%
                        }%
                        $ \t \longleftarrow \t + 1 $\;%
                    }
                    \If{$ \u = \z $}{%
                        \tcp{it is not possible to determine the proper value of $ a_j $}%
                        \KwRet{Error!}%
                    }
                    \ElseIf(\tcp*[h]{$ a_j = 1 $}){$ \u > \z $}{%
                        $ \v \longleftarrow \v + {\G}_{\j} $ \tcp*{$ v_j = {\G}_{\j} $}%
                        $ {\a}_{\j} \longleftarrow 1 $\tcp*{set the $ \j $-th element of $ \a $ as $ 1 $}%
                    }\Else{%
                        \tcp{$ a_j = 0 $}%
                    }
                }%
                \KwRet{\v, \a}
            }
        \end{algorithm}%
        \ULforem%

\newpage

    Finally, in \algref{decode-errors}, we fully decode the word $ x = a \cdot G + e $ by calling
    the \algref{decode-error-level} starting with the highest-degree coefficient. More precisely,
    first we reverse sort the keys of the dictionary of redundancy relations computed in \algref{decoding-set}.
    As the keys of the latter are the degrees of the coefficients, the first element into the list $ L $ is the
    highest-degree, and the last element of $ L $ is the lowest-degree. Second we call the \algref{decode-error-level}
    for each degree in $ L $, and we subtract the resulting word from the previous one. Indeed, we call
    the \algref{decode-error-level} with word $ v $, which is
    \begin{itemize}
        \itemsep0em
        \item $ x $ before the first call of the \algref{decode-error-level};
        \item $ x - x^\prime $ after the first call of the \algref{decode-error-level},
            where $ x^\prime $ is the result of the \algref{decode-error-level}, that is,
            $ x^\prime = \sum a_i v_i $, for each coefficient $ a_i $ of highest-degree.
    \end{itemize}
    After the loop, we will have that $ v = a_0 v_0 + e $. Hence, in order to decode $ a_0 $,
    we count the number of bits equal to $ 1 $ in $ v $. If the number of $ 1 $ is greater then $ \frac{n}{2} $,
    then $ a_0 $ is equal to $ 1 $, while if the number of $ 1 $ is lower than $ \frac{n}{2} $, then $ a_0 $ is equal to $ 0 $.

\FloatBarrier
        \normalem
        \begin{algorithm}[tbh!]
            \label{alg:decode-errors}
            \caption{Decode}
            \small
            \SetAlgoNoLine
            \SetAlgoNoEnd
            \SetInd{0.5em}{0.5em}
            \SetFillComment
            \DontPrintSemicolon

            \SetStartEndCondition{ $ \left(\right. $}{$ \left. \right) $}{$ \left. \right) $}\SetAlgoBlockMarkers{}{\}}%
            \SetKwIF{If}{ElseIf}{Else}{if}{ \{}{elif}{else \{}{}%
            \SetKwFor{For}{for}{ \{}{}%
            \SetKwFor{While}{while}{ \{}{}%
            \SetKwProg{Fn}{Function}{ \{}{}
            \AlgoDisplayBlockMarkers%

            \SetKwData{G}{G}
            \SetKwData{a}{a}
            \SetKwData{ap}{$ a^\prime $}
            \SetKwData{v}{v}
            \SetKwData{vp}{$ v^\prime $}
            \SetKwData{e}{e}
            \SetKwData{x}{x}
            \SetKwData{k}{k}
            \SetKwData{n}{n}
            \SetKwData{i}{i}
            \SetKwData{l}{r}
            \SetKwData{u}{u}
            \SetKwData{R}{R}
            \SetKwData{L}{L}
            \SetKwFunction{Decode}{Decode}
            \SetKwFunction{DecodeRelation}{Decode-Level-R}

            \Fn{\Decode{\x, \G, \R}}{%
                \KwData{$ \a \cdot \G + \e = \x \in {\F_2}^{\n} $, $ \G \in {\F_2}^{\k \times \n} $, $ \R $, where $ \x $ is the vector
                    to decode, $ \G $ is the generator matrix of the HL-code, and $ \R $ is the dictionary of the redundancy relations
                    for the coefficients $ a_j $ of degree $ \l $ as in \DecodingSet}
                \KwResult{$ \a \cdot \G $}
                \BlankLine

                $ \k \longleftarrow $ get the number of rows of $ \G $\;%
                $ \n \longleftarrow $ get the number of columns of $ \G $\;%
                $ \a \longleftarrow \mathbf{0} \in {\F_2}^{k} $\;%
                $ \v \longleftarrow \x $\;%
                $ \L \longleftarrow $ reverse sort the keys in $ \R $\;%
                \ForAll{$ \l $ in $ \L $}{%
                    $ (\vp, \ap) \longleftarrow \DecodeRelation{\R, \l, \v, \G} $\;%
                    $ \v \longleftarrow \v - \vp $\tcp*{$ \v - \sum a_j v_j $}%
                    $ \a \longleftarrow \a + \ap $\;%
                }
                \tcp{$ \v = a_0 \v_0 + \e $}%
                $ \u \longleftarrow $ count how many components of $ \v $ are equal to $ 1 $\;%
                \If{$ \u = \frac{\n}{2} $}{%
                    \tcp{it is not possible to determine the proper value of $ a_0 $}%
                    \KwRet{Error!}%
                }
                \ElseIf(\tcp*[h]{$ a_0 = 1 $}){$ \u > \frac{\n}{2} $}{%
                    $ \v \longleftarrow \v - {\G}_0 $ \tcp*{$ v_0 = {\G}_0 $}%
                    $ {\a}_0 \longleftarrow 1 $\tcp*{set the first element of $ \a $ as $ 1 $}%
                }\Else{%
                    \tcp{$ a_0 = 0 $}%
                }%
                \tcp{now $ \v = \e $ and $ \x - \v = \a \cdot \G $}
                \KwRet{$ \x - \v $, $ \a $}
            }
        \end{algorithm}%
        \ULforem%

    To summarize, in order to speedup the whole process, we divide our algorithm in two steps.
    First, we
    \begin{itemize}
        \itemsep0em
        \item Compute the partial generator matrix using the \algref{build-rm-matrix}, and we save it into a file;
        \item Compute the redundancy relations for $ a_j $, with $ 1 \le j < k - \frac{1}{2} \binom{m}{m / 2} $, and we
            save them into a file along with the cache.
    \end{itemize}
    Once we saved the partial generator matrix and the redundancy relations for this matrix, we
    use the McEliece protocol to encrypt/decrypt a message. More precisely, we define the
    \protref{key-gen}, \protref{enc}, and \protref{dec}.

    \normalem

   \FloatBarrier
    \begin{protocol}
        \label{prot:key-gen}
        \caption{Key-Gen}
        \KwIn{The parameter $ m \in \N $ of the HL-code}
        \KwResult{Return the private-key and the public-key}
        \begin{enumerate}
            \item
            Retrieve the partial generator matrix for $ m $ from the file, and get its dimension $ k \times n $;

            \item
            Compute a random complement-free set $ Y $ with \algref{build-y-set}, and compute the last rows of
            the generator matrix $ G $ with \algref{build-hl-matrix};

            \item
            Retrieve the redundancy relations for $ m $ from the file, and compute the remaining redundancy relations
            with \algref{decoding-set} for $ j_{\text{start}} = k - \frac{1}{2} \binom{m}{m / 2} $,
            and $ j_{\text{end}} = k - 1 $. We call $ R $ the complete dictionary of all the redundancy relations;

            \item
            Compute an invertible matrix $ S \in {\F_2}^{k \times k} $, and its inverse $ S^{-1} $;

            \item
            Compute a permutation $ \rho \in \func{Sym}{\brackets{1, \ldots, n}} $, and its inverse $ \rho^{-1} $;
        \end{enumerate}
        Return the private-key $ (S^{-1}, \rho^{-1}, G, R) $ and the public-key
        $ G^\prime = \func{\rho}{S \cdot G} $, where $ \func{\rho}{S \cdot G} $ permutes the columns
        of $ S \cdot G $.
    \end{protocol}
    \begin{protocol}\label{prot:enc}
        \caption{Encrypt}
        \KwIn{The public-key $ G^\prime \in {\F_2}^{k \times n} $,
            the parameter $ m $ of the code, and a message $ m \in {\F_2}^k $}
        \KwResult{Return the encrypted message}
        \begin{enumerate}
            \item
            Compute the minimum distance $ d = 2^{\frac{m}{2}} $, and $ t = \floor{\frac{d - 1}{2}} $;

            \item
            Compute a random word $ e \in {\F_2}^{n} $ such that $ \wt{e} = t $;
        \end{enumerate}
        Return the encrypted message $ c = m \cdot G^\prime + e $.
    \end{protocol}
\FloatBarrier
    \begin{protocol}\label{prot:dec}
        \caption{Decrypt}
        \KwIn{The private-key $ (S^{-1}, \rho^{-1}, G, R) $,
            the parameter $ m $ of the code, and the encrypted message $ m \cdot G^\prime + e = c \in {\F_2}^n $}
        \KwResult{Retrieve the message $ m $ from the encrypted message $ c $}
        Decode the word $ \func{\rho^{-1}}{c} $ with \algref{decode-errors},
        and get $ m \cdot S $;
        Return the message $ m = x \cdot S^{-1} $.
    \end{protocol}
    \ULforem%

    In the following, we present our results implementing the above algorithms and protocols
    in Python V3.6 using the math library SageMath \footnote{\url{https://www.sagemath.org/}} V9.0.
    The hardware of the machine used for these implementations is in \tabref{laptop}.
    We benchmark our algorithms with the function \textbf{perf\_counter} (which measure the WALL-time)
    of the library \textbf{time} of Python 3.6.

  \FloatBarrier
    \begin{table}[!htpb]
        \footnotesize
        \caption{Hardware}
        \label{tab:laptop}
        \centering
        \begin{tabular}{ll}
            \toprule
            SSD & SAMSUNG 850 EVO - 256 GB \\
            \midrule
            RAM & 12 GB - 1600 MHz\\
            \midrule
            Processor & i5-5200U 2.2 GHz \\
            \bottomrule
        \end{tabular}
    \end{table}

    As we implemented these algorithms with Python, we further optimize the code
    by representing each redundancy relation as an integer of $ n = 2^m $ bits.
    More precisely, as $ x = a \cdot G + e = \roundbrack{x_0, x_1, \ldots, x_{n - 1}} $, we
    associates each redundancy relation $ \underset{e}{\Delta}\  x_i = \bigoplus x_j $ to an integer such that the
    $ j $-th component is $ 1 $.
    For instance, if $ n = 16 $ and the redundancy relation is $ x_0 \oplus x_1 \oplus x_4 \oplus x_5 $
    (see the first redundancy relation of $ a_6 $ in \eqref{eq:rec-rel-1}),
    then $ x_0 \oplus x_1 \oplus x_4 \oplus x_5 = {(1, 1, 0, 0, 1, 1, 0, 0, 0, 0, 0, 0, 0, 0, 0, 0)}_2 = 52224 $.
    By using these representation, we pre-compute the bit-arrays for each components of $ x $ and we use them
    to optimize \algref{aj-redundancy-relations} and \algref{decode-error-level}. More precisely,
    in \algref{aj-redundancy-relations} we compose the redundancy relation as an bitwise-OR of the
    corresponding bit-arrays. For instance, $ x_0 \oplus x_1 $ is the bitwise-OR of the bit-array
    of $ x_0 $ (e.g., $ x_0 = {(1, 0, 0, 0, 0, 0, 0, 0, 0, 0, 0, 0, 0, 0, 0, 0)}_2 = 32768 $), and
    the bit-array of $ x_1 $ (e.g., $ x_1 = {(0, 1, 0, 0, 0, 0, 0, 0, 0, 0, 0, 0, 0, 0, 0, 0)}_2 = 16384 $),
    that is, $ x_0 \oplus x_1 = 32768 \text{ OR } 16384 = 49152 $.
    In this way, the set of redundancy relations
    for each $ a_j $ is a set of integers. Moreover, in \algref{decode-error-level} we also represent the word $ x $
    as an integer and the evaluation of each redundancy relations correspond to count the number of bits equals to
    $ 1 $ in $ \roundbrack{\underset{e}{\Delta}\  x_i} \ \&\  x $,
    where $ (\&) $ is the bitwise-AND between the respective integer representations. For instance, evaluate
    the redundancy relation $ x_0 \oplus x_1 \oplus x_4 \oplus x_5 $ is equal to count the number of bits
    equals to $ 1 $ into the integer $ 52224 \ \&\  x $. Indeed, as $ 1 \oplus 1 = 0 $,
    the redundancy relation is equal to $ 1 $ if and only if the number of $ 1 $ of the involved
    components is odd. Hence, if the number of ones in $ \roundbrack{\underset{e}{\Delta}\  x_i} \ \&\ x $ is odd,
    then $ \underset{e}{\Delta}\  x_i $ is equal to $ 1 $.

\begin{figure}[!htbp]
        \centering
        \begin{minipage}[b]{0.5\linewidth}
            \includegraphics[width=\linewidth]{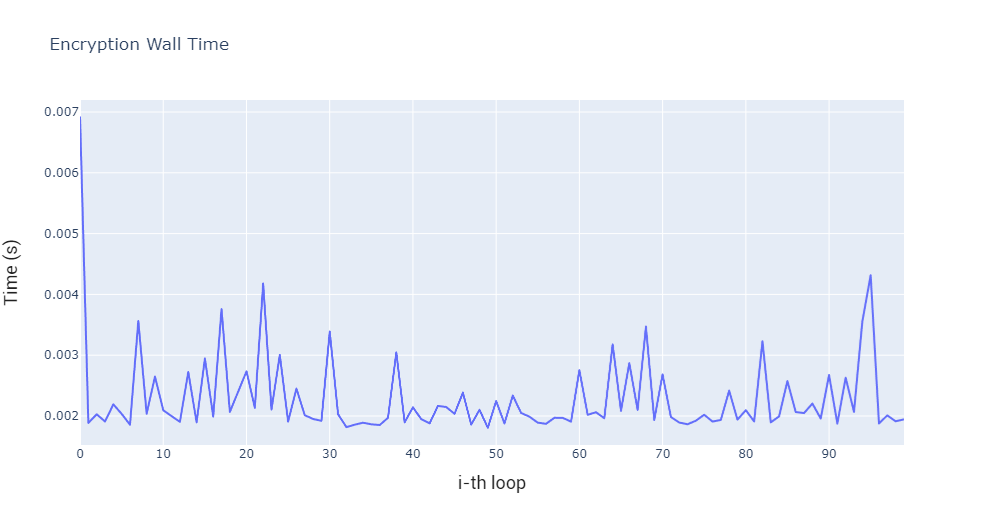}
            \caption{{\scriptsize Encryption times for an HL-code of parameter $ m = 10 $}}
            \label{fig:enc10}
        \end{minipage}
        \begin{minipage}[b]{0.5\linewidth}
           	\includegraphics[width=\linewidth]{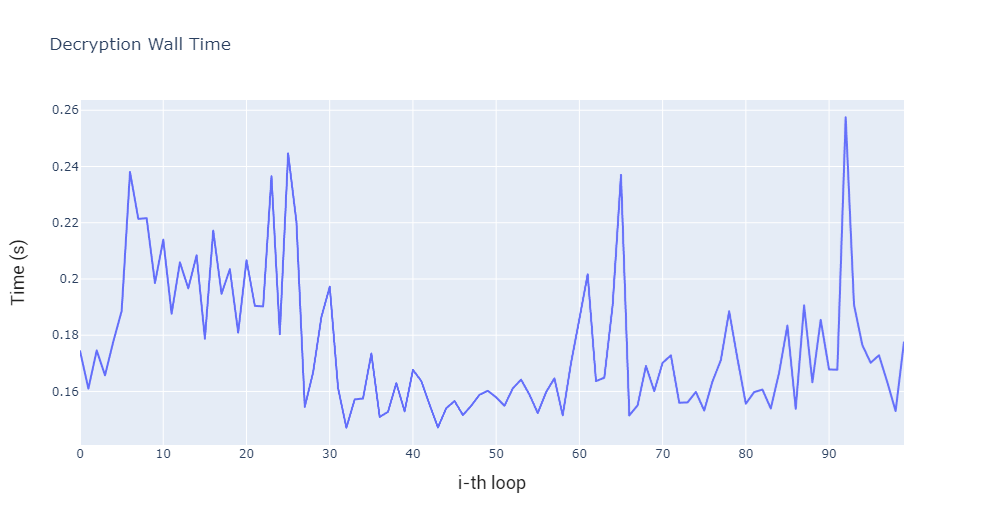}
           	\caption{{\scriptsize Decryption times for an HL-code of parameter $ m = 10 $}}
           	\label{fig:dec10}
        \end{minipage}\hfill
        \begin{minipage}[b]{0.5\linewidth}
            \includegraphics[width=\linewidth]{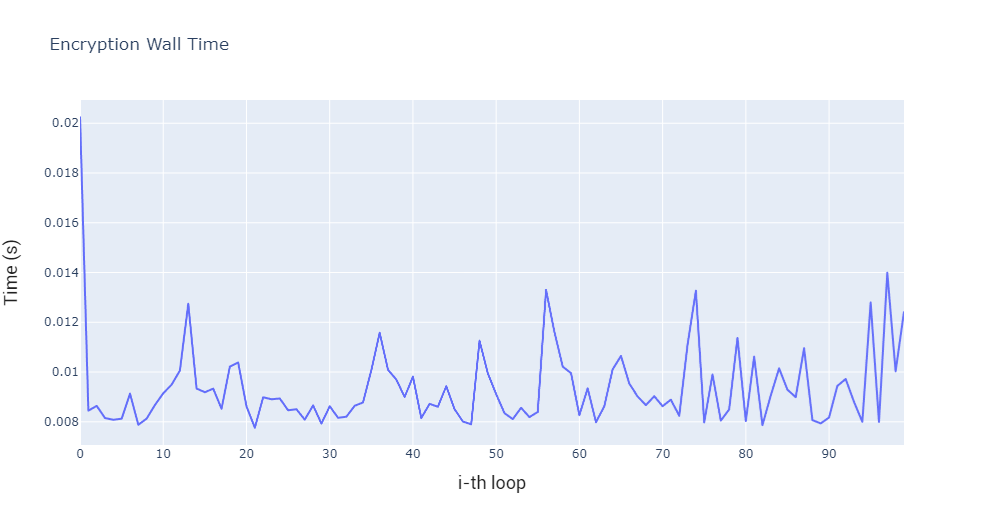}
            \caption{{\scriptsize Encryption times for an HL-code of parameter $ m = 12 $}}
            \label{fig:enc12}
        \end{minipage}
        \begin{minipage}[b]{0.5\linewidth}
            \includegraphics[width=\linewidth]{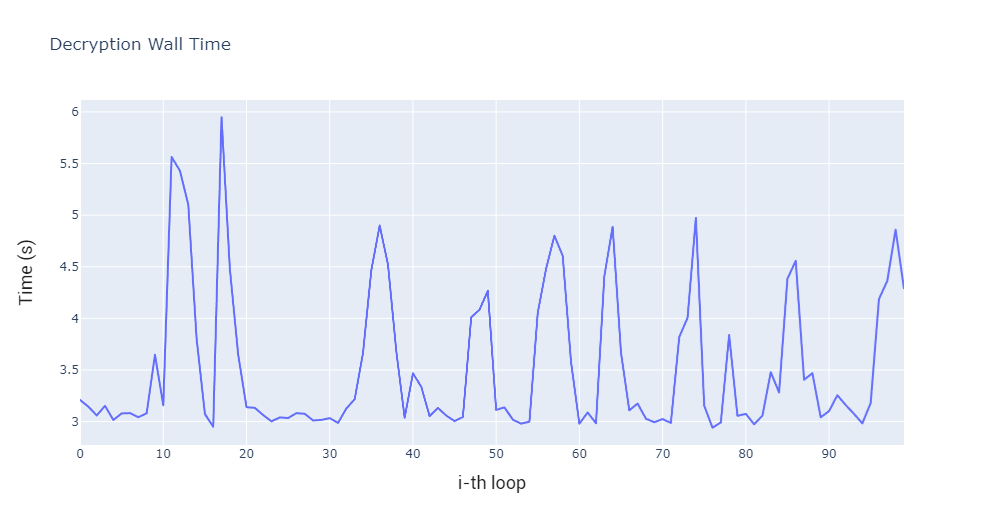}
            \caption{{\scriptsize Decryption times for an HL-code of parameter $ m = 12 $}}
            \label{fig:dec12}
        \end{minipage}
        \caption{WALL-time for an HL-code of parameters $ m = 10, 12 $}
    \end{figure}

    In \figref{enc10} and in \figref{dec10} we plot the encryption times and the decoding times
    (in seconds), respectively, by using our algorithms for a random HL-code with parameter $ m = 10 $.

     \FloatBarrier

    In \figref{enc12} and in \figref{dec12} we plot the encryption times and the decoding times
    (in seconds), respectively, by using our algorithms for a random HL-code with parameter $ m = 12 $.

    In \tabref{plot-mean-time} we show the mean time for the encryption and the decryption,
    and the time for the key-generation phase. We remark that in the \protref{key-gen} we retrieve
    the partial generator matrix (computed with \algref{build-rm-matrix}) from the disk. In particular,
    for $ m = 12 $, we have that the \protref{key-gen} required $ 6.59 $ seconds including $ 3.2 $ seconds
    to read the matrix from the disk.

   \FloatBarrier
    \begin{table}[!htbp]
    	\footnotesize
    	\caption{Mean time for the protocol}
    	\label{tab:plot-mean-time}
    	\centering
    	\begin{tabular}{llll}
    		\toprule
    		m & Key-Gen (s) & Mean encryption time (s) & Mean decryption time (s) \\
    		\midrule
    		10 & 0.3297 & 0.0023 & 0.1758 \\
    		\midrule
    		12 & 6.5884 & 0.0094 & 3.5341 \\
    		\bottomrule
    	\end{tabular}
    \end{table}

    \FloatBarrier

    \section{Conclusion}

    In this paper we have shown that there exists a decoding algorithm for HL-codes, which can be implemented efficiently. Further, we implemented the DHH-cryptosystem introduced in \cite{domosi2019cryptographic} and showed that its running time is in acceptable range on an every-day-used device today.

    %
    %
    %
    %
    %
    %
    %
    %

    \bibliography{biblio.bib}{}
    \bibliographystyle{unsrt}

\end{document}